\documentclass[12pts,aps,twocolumn,prd,showpacs,nofootinbib,preprintnumbers]{revtex4-1}
\usepackage{bm}
\usepackage{latexsym}
\usepackage{natbib}
\usepackage{url}
\usepackage{dcolumn}
\usepackage{color}
\usepackage{amsfonts}
\usepackage{amssymb}
\usepackage{amsmath}
\usepackage{mathtools}
\usepackage{graphicx,epsfig}
\usepackage{psfrag}
\usepackage{subfigure}
\usepackage{natbib}
\usepackage{array, makecell}
\usepackage[latin1]{inputenc}
\usepackage{graphicx}
\graphicspath{{images/}}

\begin{document}

\title{Unified $f(R)$ gravity at local scales}

\author{ Vipin Kumar Sharma}
\email[]{\textcolor[rgb]{1.00,0.00,0.00}{vipinastrophysics@gmail.com}}
\author{Murli Manohar Verma}
\email[]{\textcolor[rgb]{1.00,0.00,0.00}{sunilmmv@yahoo.com}, \textcolor[rgb]{1.00,0.00,0.00}
{murli.manohar.verma@cern.ch}}

\affiliation{$^{*, \dagger}$Department of Physics, University of Lucknow, Lucknow 226 007, India \\
$^{\dagger}$Theoretical Physics Department, CERN, CH-1211 Geneva 23, Switzerland  }
\date{\today}
\begin{abstract}
We explore the shifted $f(R) (\propto R^{1+\delta})$ model with ${\delta}$ as a distinguishing physical parameter for the  study of constraints  at local scales. The corresponding  dynamics confronted with different geodesics (null and non-null) along with its conformal analogue is investigated. For null geodesics, we discuss the light deflection angle, whereas for non-null geodesics under the  weak field limit, we investigate the perihelion advance of the Mercury orbit in $f(R)$ Schwarzschild background, respectively. The extent of an additional force, appearing  for non-null geodesics, depends on $\delta$.  Such phenomenological investigations allow us to strictly constrain $\delta$ to be approximately $\mathcal{O}(10^{-6})$ with a difference of unity in orders at galactic and planetary scales and seems to provide a unique $f(R)$ at local scales.
Further, at late cosmic time, we analyse the constraint on $\delta$ via the bare scalar self-interaction Einstein frame potential to provide a null test of dark energy.
We constrain the deviation parameter, $\mid\delta\mid$ to  $(\approx 0.6)$ which is in a close agreement with the results obtained through various observations in the Jordan frame by several  authors. Our results suggest that the present form of model is suitable for the alternate explanation of dark matter-like effects at local scales, whereas at large scales the deviations grow higher and must be addressed in terms of the accelerated background.
\end{abstract}
\maketitle{}

\section{\label{1}Introduction}
Precision cosmology is  imperfect without the precise physical observations of systems at different scales. For precise observations, the physical system must be completely known. Unfortunately, the major portion of our observable universe is not completely known to us in the Einstein's General Relativity (GR) framework. Also, the Einstein's gravity theory with positive cosmological constant ($\Lambda$) based on GR supported by observations still can not be regarded as an ultimate gravity theory because of the recent stronger than before Planck's observation of an accelerating universe ( having the shifted $w$ from -1) and its inability to reconcile with the quantum theory completely and also more importantly, the fundamental properties of dark sector (dark energy and cold dark matter) is still unknown besides its distribution. Thus, the observational evidences at different redshifts nurture the concept of modified gravity theory instead of the Einstein's physical gravity theory of GR with $\Lambda$ (positive) and cold dark matter (CDM) \textcolor[rgb]{0.00,0.00,1.00}{\cite{b1,b2,b3,b4,b5,b6,b7,b8, b08}}. Actually, the investigation of deviation in the Einstein's GR theory has become an active area of research under the study of precision concordance cosmology. Therefore, it is the need of precision concordance cosmology to opt for an alternative theory of gravity.

Brans and Dicke first explored the scalar-tensor approach of gravity theory as an alternative to the Einstein-Hilbert (E-H) gravity theory \textcolor[rgb]{0.00,0.00,1.00}{\cite{b9}} and thereafter H. A. Buchdahl, A. A. Starobinsky and others carried out the study of higher order Lagrangian or non-linear Lagrangian for exploring the cosmic evolution and also for the phenomenological explanation of the major observational issues at different cosmological redshifts of the universe \textcolor[rgb]{0.00,0.00,1.00}{\cite{b9,b09,b10,b11,b12,b13,b14,b15,b16,b17,b18,b19,b20,b21,b22,b23,b24,
 b25,b26,b27,b28,b29,b30,b31,b32,b032}}.
The main implicit powers of the alternate gravity theory can be understand in two ways, on one hand it can address  dark sector (dark matter and dark energy) with the different particulate degrees of freedom (see \textcolor[rgb]{0.00,0.00,1.00}{\cite{b20}}) and on the other hand it can precisely address the deviation in the existing gravity theory i.e. Einstein's GR theory, which at present is very active nowadays in literatures \textcolor[rgb]{0.00,0.00,1.00}{\cite{b45,b35,b33,b34,b80}}. Such alternate gravity theory may also provide a possible way to study the unified theory of dark matter and dark energy \textcolor[rgb]{0.00,0.00,1.00}{\cite{b033}} .

In this direction, motivated by the dynamical modelling of massive test masses in $f(R)$ gravity model by B$\ddot{o}$hmeret al.  \textcolor[rgb]{0.00,0.00,1.00}{\cite{b36}} and also by our recently investigated power-law $f(R)$ model at low redshifts \textcolor[rgb]{0.00,0.00,1.00}{\cite{b33,b34,b35}} and by others \textcolor[rgb]{0.00,0.00,1.00}{\cite{b37,b38,b39,b40}} for the clustered dark matter explanation, we further analyse here the precise and strict phenomenological constraints on the model parameter $\delta$ in $R\rightarrow R^{1+\delta}$. That is, we investigate here the constraints corresponding to different geodesics (null and non-null) along with its conformal analogue at the galactic and planetary scales under an idealistic consideration with the $f(R)$ modified potential explored in \textcolor[rgb]{0.00,0.00,1.00}{\cite{b38,b39,b40}}. Moreover, our recent combined analysis of light deflection profiles via rotation curves for typical spiral massive galaxies may in principle also diagnose the present $f(R)$ model from the other profile of standard dark matter model of galaxy (say Pseudo-isothermal model) in terms of lensing angle \textcolor[rgb]{0.00,0.00,1.00}{\cite{b34}}. With such motivations and also from the motivation drawn from the physical form of deviation introduced in $f(R)$ model by B$\ddot{o}$hmer et al.  \textcolor[rgb]{0.00,0.00,1.00}{\cite{b36}}, we look for $\delta$ as a distinguishing parameter for investigating the precise constraints through the study of motions of test masses under the stable orbits in different geodesics (null and non-null) along with its conformal analogue. Although such power-law $f(R)$ modification was initially carried out for the study of early-time cosmic evolution, late-time cosmic evolution and also for the galactic dynamic explanation  \textcolor[rgb]{0.00,0.00,1.00}{\cite{b41,b42,b43}} and \textcolor[rgb]{0.00,0.00,1.00}{\cite{b36,b37,b38,b39,b40}}.

From the phenomenological point of view, it seems that $f(R)$ formulation have spaces for the explanation of most abundant dark sector of the energy budget of our universe with the standard matter. From the mathematical point of view and motivated by the earlier Jordan-Brans-Dicke Lagrangian
\textcolor[rgb]{0.00,0.00,1.00}{\cite{b9}}, the $f(R)$ action formulation can be recast into the scalar-tensor form so that one can clearly see the appearance of an extra scalar degree of freedom (scalar fields) which occupy the spaces of dark sectors and unlike some gravity theories in which we have an ad hoc scalar fields \textcolor[rgb]{0.00,0.00,1.00}{\cite{b44}}.
Such an extra $f(R)$ scalar degree of freedom has  non-minimal coupling with the spacetime geometry. Thus, in alternate gravity theories, gravitational interaction can also be mediated via the scalar degrees of freedom (scalaron) together with the spacetime metric tensor ($g_{\mu\nu}$). $f(R)$ formulation has also two characteristic length scales, of which one is the Schwarzschild length scale and the other is the $f(R)$ characteristic length scale which has an advantage to provide a screening mechanism  \textcolor[rgb]{0.00,0.00,1.00}{\cite{b45}} corresponding to the extra force in high density environments.

For  investigating the physical observable effects of such extra $f(R)$ scalar degrees of freedom alone
(scalaron as scalar field particle), it is preferred in literature  that extra scalar degree of freedom should be coupled minimally with the Einstein-Hilbert spacetime geometry ($R$) in the action integral so that we can trace its dynamical influence via solving the second order field equations instead of fourth order field equations. This can be possible through the conformal analysis of spacetime metric tensor \textcolor[rgb]{0.00,0.00,1.00}{\cite{b44}}.
Thus, we have two $f(R)$ gravity actions before and after the conformal transformation i.e., Jordan frame (because the dynamical equations resembles the Jordan-Brans-Dicke gravity theory) and conformal Einstien frame (because the dynamical equations resembles with the Einstein-Hilbert gravity theory). Also, pertaining to the physical observations at different scales, which frame  one should prefer is still under debate and several research works have been reported from the early investigations \textcolor[rgb]{0.00,0.00,1.00}{\cite{b46,b47,b48,b49,b50,b51}}, till recent analyses
\textcolor[rgb]{0.00,0.00,1.00}{\cite{b52,b53,b54,b55,b56,b57,b58}}.
Hence, the well known issue in literature regarding the choice of true physical frame is actually a long standing i.e. which spacetime metric tensor among the conformally related  ($g_{\mu\nu}$ or $\tilde{g}_{\mu\nu}$ ) should one may choose for representing the geometry of spacetime pertaining to the physical observations. Some scientists claim that both frames are equivalent whereas others claim opposite ( in references \textcolor[rgb]{0.00,0.00,1.00}{\cite{b55,b56,b57,b58}}). The issue can only be settled down through the physical observations of the system.

Several authors had explored this issue before without any firm discussions regarding the most profound physical observations confronted with different geodesics (null and non-null) along with its conformal analogue under the shifted ($R\rightarrow R^{1+\delta}$) Einstein-Hilbert theory at local scales. Thus, along with discussing the conformally related  different geodesics, we also explore the nature of an additional force in terms of $f(R)$ model parameter $\delta$ and further envisage the shifted parameter ($\delta$) as a distinguishing tool  for the study of  motions at different geodesics to impose the strict constraints on the $f(R)$ model via observations.

We also acquired the motivation to explore the bare effects of $f(R)$ modification at different  scales. Like, from the important analysis to the limit to GR in $f(R)$ theory of gravity by G. J. Olmo \textcolor[rgb]{0.00,0.00,1.00}{\cite{b59}}, from the study of $f(R)$ model constraints at different scale by W. Hu and I. Sawicki \textcolor[rgb]{0.00,0.00,1.00}{\cite{b27}}, from the recent and early investigation of the Yukawa-like potential in $f(R)$ theory  \textcolor[rgb]{0.00,0.00,1.00}{\cite{b60,b61}}, from the study of power-law $f(R)$ gravity theory \textcolor[rgb]{0.00,0.00,1.00}{\cite{b37,b38,b39,b40}}, from the recent investigation of an uncertainty in the GR prediction for the Mercury orbit under the current (MESSENGER) and planned (the European-Japanese BepiColombo) missions \textcolor[rgb]{0.00,0.00,1.00}{\cite{b62,b63,b64,b65,b66}}, from the analysis of Yukawa type potential in the Schwarzschild-like background for perihelion precession of planets to address the gauge bosons as a possible candidate of fuzzy dark matter\textcolor[rgb]{0.00,0.00,1.00}{\cite{b066}} and from the relativistic effects and dark matter in the Solar system through the observations of planets and spacecraft \textcolor[rgb]{0.00,0.00,1.00}{\cite{b067}}.

In the present work, we study the deviations in GR in the form $R^{1+\delta}$ (with $\delta$ being the  dimensionless physical observable quantity) and instead of investigating the Yukawa-like correction potential as others did \textcolor[rgb]{0.00,0.00,1.00}{\cite{b60,b61}}, we explore the bare modified $f(R)$ effects in the form of $\delta$ for different observations without any versions of dark sector. Such form can be modelled in the $f(R)$ gravity framework. Also, since physical observations of the system is confronted with different geodesics, we obtain its conformal analogue and investigate the precise deviations at different scales. For instance, we phenomenologically study the geodesics (null and non-null) along with its conformal analogue at local scales under an idealistic considerations and obtain the strict constraints on  $\delta$ at galactic and planetary scales, respectively. Such constraints are strict in the sense that for orbital motions at local scales, $\delta$ can be physically represented as $v_{orbital}^2/c^2$ with known $v_{orbital}$ \textcolor[rgb]{0.00,0.00,1.00}{\cite{b36}}.  Further, we explore the extent and nature of an additional force on the test mass due to  the shifted $f(R)$ modification parameterized by $\delta$. Next, we analyse the late-time cosmic constraint via the bare $f(R)$ potential of Einstein-frame to provide a null test of dark energy, which is also investigated by S. Capozziello under different cosmic observations \textcolor[rgb]{0.00,0.00,1.00}{\cite{b77}} in jordan frame. The work is followed by the general discussion of conformally transformed $f(R)$ gravity action and their field equations.
The present paper is organised in different sections as follows. In Section \textcolor[rgb]{1.00,0.00,0.00}{(II)},
we mathematically discuss the Jordan and conformal Einstein frames dynamical $f(R)$ field equations.
In Section \textcolor[rgb]{1.00,0.00,0.00}{(III)}, we  discuss the parametrization for geodesics
and phenomenologically investigate the null geodesics for the conformally related light deflection angles due to massive spiral galaxies as lens.
In Section \textcolor[rgb]{1.00,0.00,0.00}{(IV)}, we discuss the non-null geodesics and investigate its Newtonian limit for the explanation of additional force in $f(R)$ theory. Further, we explore it
phenomenologically for the conformally related orbital precession of planet (Mercury) at the solar system scale and analyse and diagnose the tight constraint via specifying $\delta$ as $v^2_{orbital}/c^2$.
Next, we  discuss the cosmic late-time constraint on the $\delta$ in Section \textcolor[rgb]{1.00,0.00,0.00}{(V)} and end with the discussion and concluding remarks of results on the constraints in Section \textcolor[rgb]{1.00,0.00,0.00}{(VI)}.

Throughout the paper, we follow the signature of the spacetime metric as ($-$, +, +, +) and
indices $\mu$ (or $\nu$) = (0, 1, 2, 3).

\section{\label{2}$f(R)$ gravity and conformally related field equations}

Lovelock theorem suggests possible modifcation to the gravity theory. According to which, Einstein's gravity theory of GR is
unique theory of gravity, if we demand \textcolor[rgb]{0.00,0.00,1.00}{\cite{b67}}:
(i) the metric tensor ($g_{\mu\nu}$) is the only field (ii) invariance under diffeomorphism (iii) the
equation of motion is second order and (iv) to work with four dimensions.

Any violation of one of these assumptions will provide an alternative or modified gravity theory.
$f(R)$ gravity theory focus on the metric tensor and equations of motion.The metric tensor
($g_{\mu\nu}$) is not the only field (but also scalar field) and also the equations of motions are not second order (but becomes forth-order) in $f(R)$ gravity theory.\\

Consider the 4-dimensional action integral having the gravity Lagrangian density possessing  some generic function of the Ricci scalar $f(R)$ with the usual standard matter action in units of c=$\hbar$=1 as,
 \begin{eqnarray}
\mathcal{A}= \frac{1}{2}\int \sqrt{-g} \left[\frac{1}{8\pi G_N}f(R)\right]d^{4}x+ \mathcal{A}_m(g_{\mu\nu}, \Psi_m) \label{a1},\end{eqnarray}
where $g$ is the determinant of the metric tensor $g_{\mu\nu}$, $8\pi G_N$ is the
Einstein's gravitational constant with $G_N$ the Newtonian gravitational constant, and $\mathcal{A}_m$ is the action of the standard matter part with matter field $\Psi_m$.
We can also write the gravity Lagrangian density of \textcolor[rgb]{0.00,0.00,1.00}{(\ref{a1})} similar to the Jordan-Brans-Dicke gravity theory as,
\begin{eqnarray}
\mathcal{A}_{J}= \int \sqrt{-g} \left[\frac{1}{2\kappa^2}R F(R)- U\right]d^{4}x \label{a2},\end{eqnarray}
where $U=\frac{RF(R)-f(R)}{2\kappa^2}$ with $F(R)=f_{R}=\frac{\partial f}{\partial R}$ and
$\kappa^2=8\pi G=M_{pl}^{-2}$.

In the action written in the form equation \textcolor[rgb]{0.00,0.00,1.00}{(\ref{a2})}, we can see that the extra scalar degrees of freedom in the form of $F(R)$ is non-minimally coupled with the geometry ($R$) and has the similar form observed in the Jordan-Brans-Dicke gravity action.
One can also explore \textcolor[rgb]{0.00,0.00,1.00}{(\ref{a1})} in Helmholtz-Jordan frame and obtainwith the same $f(R)$ action which have non-vanishing second $f(R)$ derivative.
We can perform conformal rescaling of the spacetime metric in order to explore further the dynamics of the extra scalar degrees of freedom.

Therefore, following the well established mathematical procedure of conformal transformation of scalar-tensor theory \textcolor[rgb]{0.00,0.00,1.00}{\cite{b44}}, the conformal map from the four dimensional spacetime manifold $\mathcal{M}_J$ with metric $g_{\mu\nu}$ to the four dimensional spacetime manifold $\mathcal{M}_E$ but with metric $\tilde{g}_{\mu\nu}$ i.e.,
\begin{eqnarray}
 \tilde{g}_{\mu\nu}=\Omega^2 g_{\mu\nu}=F(R) g_{\mu\nu}\label{a3},\end{eqnarray}
along with
\begin{eqnarray}
R=\Omega^2[\tilde{R} + 6 \tilde{\Box} \omega - 6 \tilde{g}^{\mu\nu} \partial_{\mu} \omega \partial_{\nu} \omega ]
\label{a4},\end{eqnarray}
whose second term vanishes when \textcolor[rgb]{0.00,0.00,1.00}{(\ref{a4})} is used in the action integral by following the Gauss' theorem of total derivative which transforms the equation
\textcolor[rgb]{0.00,0.00,1.00}{(\ref{a2})} as,
\begin{eqnarray}
 \mathcal{A}_{E}= \int d^{4}x \sqrt{-\tilde{g}}
 \left[\frac{1}{2\kappa^2}\tilde{R}-6\tilde{g}^{\mu\nu}\partial_{\mu} \omega \partial_{\nu} \omega-  V(\omega)\right]
\label{a5},\end{eqnarray}
where
\begin{eqnarray}
V(\omega)=\frac{U}{2F^2}=\frac{RF-f}{2\kappa^2 F^2} \label{a6}\end{eqnarray}
is the scalar potential term written in parametric form for general $f(R)$ model which has non-vanishing
second derivative and $\omega \equiv \ln \Omega$.\\ Here, instead of considering a new parametric
scalar field\footnote{\textcolor[rgb]{1.00,0.00,0.00}{The linear canonical action with minimally coupled
scalar field is obtained by defining a new scalar field $\kappa \phi(\equiv \sqrt{\frac{3}{2}}\ln F)$ with
$\omega= \kappa \phi \sqrt{\frac{1}{6}}$}},
we workout with the action \textcolor[rgb]{0.00,0.00,1.00}{(\ref{a5})} in its simplest form and explore its
field equations.

The field equations in this frame are obtained by varying \textcolor[rgb]{0.00,0.00,1.00}{(\ref{a5})} w.r.t
the metric tensor (${g}^{\mu\nu}$) and the scalar field ($\omega$) which are given as \textcolor[rgb]{0.00,0.00,1.00}{\cite{b68}},
\begin{eqnarray}
 \tilde {G}_{\mu\nu}= 6\left[\tilde {\nabla}_\mu \omega \tilde {\nabla}_\nu \omega-
 \frac{1}{2} \tilde {g}_{\mu\nu} (\tilde {\nabla} \omega)^2 \right]- \tilde {g}_{\mu\nu} V(\omega)
 \label{a7},\end{eqnarray}
 and
\begin{eqnarray}
\tilde{\Box} \omega=\frac{1}{6} V^\prime (\omega).\label{a8}\end{eqnarray}

Now, one can obtain the field equations by transforming equations \textcolor[rgb]{0.00,0.00,1.00}{(\ref{a7})} and \textcolor[rgb]{0.00,0.00,1.00}{(\ref{a8})} back to
without tilde notation i.e. in the Jordan frame as \textcolor[rgb]{0.00,0.00,1.00}{\cite{b68}},
\begin{eqnarray}
{G}_{\mu\nu}= \left[-2{\nabla}_\mu {\nabla}_\nu \omega+4 {\nabla}_\mu \omega {\nabla}_\nu \omega \right]-
{g}_{\mu\nu} W(\omega) \label{a12},\end{eqnarray}
where
\begin{eqnarray}
W( \omega)=\frac{f+RF}{6F} V^\prime (\omega).\label{a13}\end{eqnarray}
and
\begin{eqnarray}
{\Box} \omega=\frac{e^{2\omega}}{6} V^\prime (\omega)+2(\nabla \omega)^2 .\label{a14}\end{eqnarray}
Hence, we get the conformally related equations.

Thus, one can mathematically transform  set of equations  between the two versions of frame with $g_{\mu\nu}$ or $\tilde{g}_{\mu\nu}$.  But their physical equivalence is still under debate \textcolor[rgb]{0.00,0.00,1.00}{\cite{b47,b48,b49,b52,b53,b54,b55,b56,b57,b58}}
and can be settled down through the observations. We wish to obtain the precise constraints  through conformal analysis that suits the physical observations in both frames. \\

Therefore, we investigate the null and non-null geodesic equations at the local scales. We then explore the light deflection angle due to a typical massive galaxies and also the perihelion advance of the planet like Mercury's orbit in the shifted $f(R)$ model with parameter $\delta$. Further, we study the late-time cosmic constraints on $\delta$.

\section{\label{3}$f(R)$ conformal investigation of geodesics and phenomenology for null geodesics}
The conformal transformation generalizes the spacetime metric according to the form of $f(R)$ (see equation \textcolor[rgb]{0.00,0.00,1.00}{\ref{a3}}).
Here, we investigate the geodesic equation in different ways i.e., under the re-parametrization and under the conformal analysis for addressing the null and non null dynamics. We express its conformally shifted parameter with new parameter ($\rho$) instead of using tilde notations. The statement can be understood as
the action is re-parametrization invariant,
so it is possible to express the geodesics (which is a geometrical entity) with different parameter,
say $\lambda$. For instance, the geodesics equations in terms of proper time can be written for the material particle following the curve ($x^\gamma (\tau)$) as,
\begin{eqnarray}
\frac{d^2{x^\gamma}}{d\tau^2}+ \Gamma^{\gamma}_{\mu \nu}\frac{d{x^\mu}}{d\tau}\frac{d{x^\nu}}{d\tau}=0,\label{a15}
\end{eqnarray}
In terms of new parameter $\lambda$, we can express \textcolor[rgb]{0.00,0.00,1.00}{(\ref{a15})} as,
\begin{eqnarray}
\frac{d^2{x^\gamma}}{d\lambda^2}+\Gamma^{\gamma}_{\mu \nu}\frac{d{x^\mu}}{d\lambda}\frac{d{x^\nu}}
{d\lambda}+\frac{d{x^\gamma}}{d\lambda} \frac{d^2{\lambda}}{d\tau^2}\frac{d{\tau}}{d\lambda}=0,\label{a16}
\end{eqnarray}
Thus, we get the generalized equation for geodesics ($x^\gamma (\lambda)$).

The affine or non-affine nature of the new assigned parameter can be traced if it is expressible in terms of the old parameter ($\tau$) as $\tau\rightarrow \tau(\lambda)$ or vice versa can also be possible.
So, on switching from $x^\gamma (\tau)$ to $x^\gamma (\lambda)$, if $\tau$ and $\lambda$ has a linear relation, then only we can recover the original geodesics equations \textcolor[rgb]{0.00,0.00,1.00}{(\ref{a15})}.
Such parameters in GR literature can be assigned as affine parameters.

Now, if $x^\gamma (\lambda)$ represents the curves in $g_{\mu \nu}$, then under equation
\textcolor[rgb]{0.00,0.00,1.00}{(\ref{a3})}, the Levi-Civita connections in $\tilde{g}_{\mu \nu}$ have
the following relation with the Levi-Civita connections in ${g}_{\mu \nu}$ as,
\begin{eqnarray}
\begin{split}
{\Gamma}^{\gamma}_{\mu\nu} (\tilde{g}_{\mu \nu})=\Gamma^{\gamma}_{\mu\nu} ({g}_{\mu \nu})+
\delta^\gamma_\nu\partial_\mu(\ln\Omega)+\delta^\gamma_\mu\partial_\nu(\ln\Omega)\\-g_{\mu\nu}
\partial^\gamma(\ln\Omega).\label{a17}
\end{split}
\end{eqnarray}
Thus, the geodesic equation now gets conformally transformed under equation \textcolor[rgb]{0.00,0.00,1.00}{(\ref{a3})}
and is given as,

\begin{eqnarray}
\begin{split}
\frac{d^2{x^\gamma}}{d\lambda^2}+\Gamma^{\gamma}_{\mu \nu}\frac{d{x^\mu}}{d\lambda}\frac{d{x^\nu}}{d\lambda}+
2 \frac{d}{d\lambda}(\ln \Omega) \frac{d{x^\gamma}}{d\lambda}\\-\partial^\gamma (\ln \Omega)
g_{\mu\nu}\frac{d{x^\mu}}{d\lambda}\frac{d{x^\nu}}{d\lambda}=0
.\label{a18}
\end{split}
\end{eqnarray}

From this equation, it becomes clear that if $x^\gamma (\lambda)$ represents
the geodesics in $g_{\mu \nu}$, then it may not be a geodesic in $\tilde{g}_{\mu \nu}$. Therefore, for the curve $x^\gamma (\lambda)$ in $\tilde{g}_{\mu \nu}$, the parameter $\lambda$ must also be conformally shifted.

It is interesting that equation \textcolor[rgb]{0.00,0.00,1.00}{(\ref{a18})} resembles the generalized
geodesic equation \textcolor[rgb]{0.00,0.00,1.00}{(\ref{a16})} but under the
$f(R)$ conformal transformation and hence it is conformally modified geodesics in $f(R)$ theory.
The nature of its parameter $\lambda$ in terms of conformal shifting or affine (non-affine) as argued
for equation \textcolor[rgb]{0.00,0.00,1.00}{(\ref{a16})} can be traced for the null and non-null cases. For the null case, the last term of conformally modified geodesic equation \textcolor[rgb]{0.00,0.00,1.00}{(\ref{a18})}
vanishes because $g_{\mu\nu} u^\mu u^\nu=0$. So, equation \textcolor[rgb]{0.00,0.00,1.00}{(\ref{a18})}
gets reduced to

\begin{eqnarray}
\frac{d^2{x^\gamma}}{d\lambda^2}+\Gamma^{\gamma}_{\mu \nu}\frac{d{x^\mu}}{d\lambda}\frac{d{x^\nu}}
{d\lambda}+2 \frac{d}{d\lambda}(\ln \Omega) \frac{d{x^\gamma}}{d\lambda}=0
.\label{a19}
\end{eqnarray}

Further, to trace the conformal shift (or affine nature) of parameter $\lambda$, we reduce the equation
\textcolor[rgb]{0.00,0.00,1.00}{(\ref{a19})} as argued for equation \textcolor[rgb]{0.00,0.00,1.00}{(\ref{a16})}.

Therefore, we re-parameterize the equation \textcolor[rgb]{0.00,0.00,1.00}{(\ref{a19})}
under $\lambda\rightarrow\rho$, with $\rho$ as a new parameter (conformally shifted) and is given as,
\begin{eqnarray}
\begin{split}
\frac{d}{d\rho}\left(\frac{dx^\gamma}{d\rho}\frac{d\rho}{d\lambda}\right) \frac{d\rho}{d\lambda}+
\Gamma^{\gamma}_{\mu \nu}\frac{d{x^\mu}}{d\rho}\frac{d{x^\nu}}{d\rho} \left(\frac{d\rho}{d\lambda}
\right)^2\\+ 2\frac{d{x^\gamma}}{d\rho}\frac{d\rho}{d\lambda}\frac{d}{d\rho}(\ln \Omega) \frac{d\rho}{d\lambda}=0
.\label{a20}
\end{split}
\end{eqnarray}

We find that, if the parameter $\rho$ has a conformal relation with $\lambda$ as,
\begin{eqnarray}
\frac{d\rho}{d\lambda}=\Omega^{-2},\label{a21}
\end{eqnarray}
then under such conformal relation, equation \textcolor[rgb]{0.00,0.00,1.00}{(\ref{a20})} becomes,
\begin{eqnarray}
\frac{d^2{x^\gamma}}{d\rho^2}+ \Gamma^{\gamma}_{\mu \nu}\frac{d{x^\mu}}{d\rho}\frac{d{x^\nu}}{d\rho}=0.\label{a22}
\end{eqnarray}
Thus,  the geodesics of particles having zero rest mass remain unaffected for the conformally related spacetime equation \textcolor[rgb]{0.00,0.00,1.00}{(\ref{a3})} in $f(R)$ theory. Hence, $\rho$ can be regarded as a conformally shifted (or an affine) parameter. Therefore, the conformally transformed spacetime metric exhibits the same causal structure.

Now, for the phenomenological study of null geodesics, we explore the light deflection angle due to a typical spiral massive galaxies. Following \textcolor[rgb]{0.00,0.00,1.00}{\cite{b38,b39,b40}}, let us consider a static Schwarzschild-like metric element outside the source under the four-dimensional spacetime with $c=1$ as,
\begin{equation}
ds^{2}= -A(r)dt^{2}+B(r)dr^2+r^2( d\theta^{2}+\sin^{2}\theta d\varphi^{2}),\label{a23}
\end{equation}
where the quantities $A(r)$ and $B(r)$ act like the functions of weak gravitational potential.
The actual form of potentials can be determined via solving the vacuum $f(R)$ field equations for the function of Ricci scalar ($f(R)$) entering in the generalized Einstein-Hilbert gravity action integral.

Under the conformal transformation \textcolor[rgb]{0.00,0.00,1.00}{({\ref{a3}})},
equation \textcolor[rgb]{0.00,0.00,1.00}{(\ref{a23})} becomes,
\begin{equation}
{d\tilde{s}}^{2}= -\tilde{A}dt^{2} + \tilde{B}{d\tilde{r}}^2 +
\tilde{r}^2( d\theta^{2}+\sin^{2}\theta d\varphi^{2}).\label{a24}
\end{equation}

Now, to calculate the deflection angle in $f(R)$ gravity theory, we must know the complete form of metric element or the form of the effective potential. For the investigation of light deflection angle, we closely follow the references \textcolor[rgb]{0.00,0.00,1.00}{\cite{b38,b39}}.
It becomes clear from the discussions of \textcolor[rgb]{0.00,0.00,1.00}{\cite{b38}} that the formal expression for the total deflection angle of light ray (following null geodesics) in $f(R)$ gravity theory must remain the same under the weak field limit of Schwarzschild-like metric. The components of Schwarzschild-like metric critically depend upon the expression chosen for $f(R)$ curvature part of the Lagrangian.

Thus, for the physically motivated metric element like  equation \textcolor[rgb]{0.00,0.00,1.00}{(\ref{a23})} with $f(R)$ obeying the power-law in Ricci curvature scalar ($f(R)\propto R^{(1+\delta)}, \delta$ dimensionless constant), we have an explicit expression for the modified effective gravitational potential in the weak field limit with $A=\frac{1}{B}=1+2\Phi_{effective}(r)$
explored first by Capozziello et al.,  and further investigated by other authors
\textcolor[rgb]{0.00,0.00,1.00}{\cite{b37,b39,b40}} as well for the study of galactic dynamics. For a point-like baryonic mass $M$ in  $f(R)$ background, the effective gravitational potential is given as,

\begin{equation}
{\Phi_{effective}(r)}\approx -\frac{GM}{2r}\left[ 1+ \left(\frac{r}{r_c}\right)^{\beta}\right], \label{a25}
\end{equation}
with
\begin{equation}
\beta=\left[\dfrac{\splitdfrac{
12(\bar{\delta})^2-7(\bar{\delta})-1-
}{\sqrt{36(\bar{\delta})^4 +12(\bar{\delta})^3-
83(\bar{\delta})^2+50(\bar{\delta})+1}}}
{6(\bar{\delta})^2-4(\bar{\delta})+2}\right],\label{a26}
\end{equation}
where $\bar{\delta}\equiv (1+\delta)$ and $r_c$ is the scaling length, whose appearance is natural in $f(R)$ gravity theory due to the Noether's symmetry (see\textcolor[rgb]{0.00,0.00,1.00}{\cite{b45}} and references therein).
$\beta$ is new parameter related with the shifted $f(R)$ model parameter ($\delta$)  as in equation \textcolor[rgb]{0.00,0.00,1.00}{({\ref{a26}})} \textcolor[rgb]{0.00,0.00,1.00}{\cite{b39}}. It is clear from the equations \textcolor[rgb]{0.00,0.00,1.00}{({\ref{a25}})} and \textcolor[rgb]{0.00,0.00,1.00}{({\ref{a26}})} that for $\delta=0$ we have $\beta=0$, and the Newtonian limit is recovered.

\begin{figure}[!h]
\centering  \begin{center} \end{center}
\includegraphics[width=0.44 \textwidth,origin=c,angle=0]{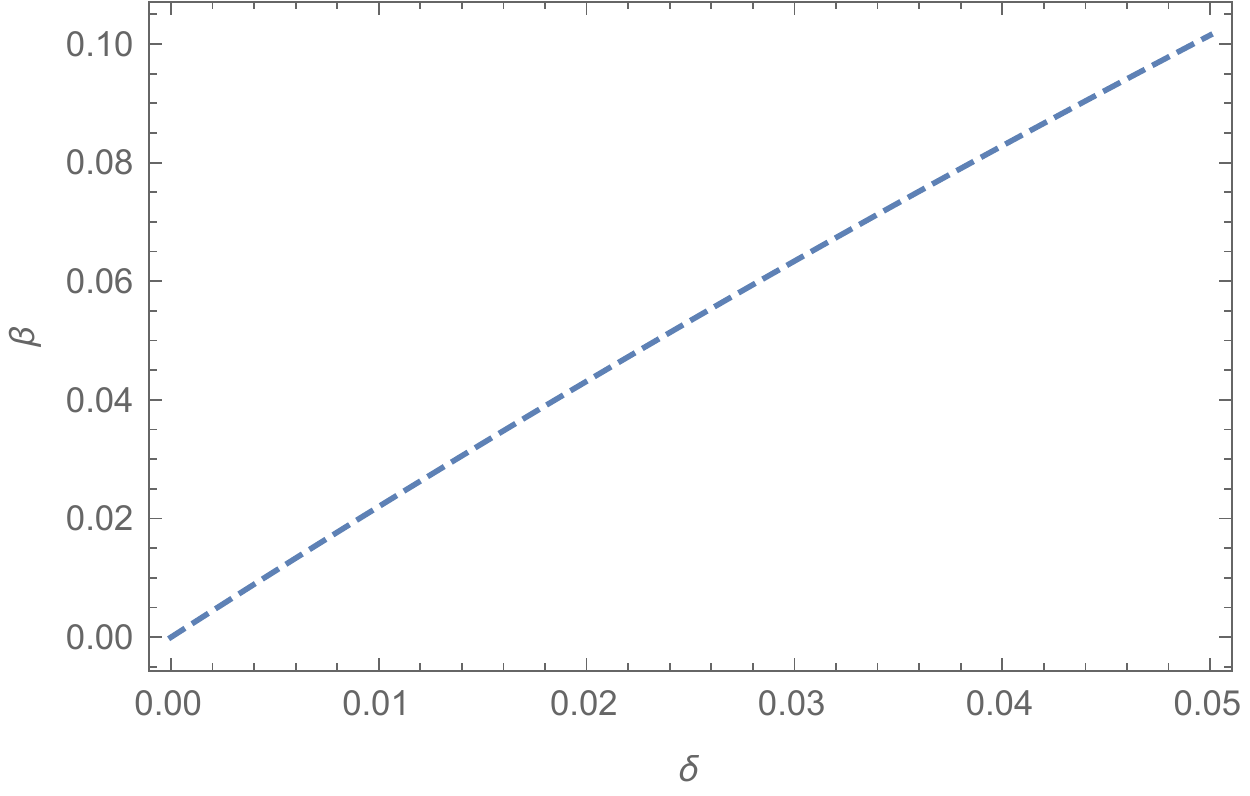}
\caption{\label{fig:p1} The curve shows the behaviour of small profile of $\delta$ and $\beta$. It is cleared from the plot that in the narrowest range,
we can consider $\mathcal{O}$($\delta$)$\approx$ $\mathcal{O}$($\beta$).}\label{f1}
\end{figure}

Furthermore, we have also recently investigated the constraint on the model parameter $\delta$ to be $\mathcal{O}(10^{-6})$ from
the combined study of flatness profile of rotation curves and the corresponding deflection angle profile in the halo of scalaron cloud (instead of dark matter) surrounding the typical massive galaxies \textcolor[rgb]{0.00,0.00,1.00}{\cite{b23,b24}}. B$\ddot{o}$ehmer et al., investigated the small value of the power-law $f(R)$  model \textcolor[rgb]{0.00,0.00,1.00}{\cite{b36}} valid for the discussion at local scale.
Therefore, it is possible to rewrite equation \textcolor[rgb]{0.00,0.00,1.00}{({\ref{a25}})} in the
narrowest range of $\delta$ and $\beta$ for the study of local scale dynamics. Since, in the narrowest range, the order of $\delta$ and $\beta$ have approximately one-to-one relation (see Fig. \textcolor[rgb]{0.00,0.00,1.00}{\ref{f1}}).
Thus, we can express equation \textcolor[rgb]{0.00,0.00,1.00}{({\ref{a25}})} in terms of $f(R)$ model
parameter $\delta$ with smaller values as,
\begin{equation}
{\Phi_{effective}(r)}\approx -\frac{GM}{2r}\left[ 1+ \left(\frac{r}{r_c}\right)^{\delta}\right]. \label{a27}
\end{equation}
Clearly, for $\delta=0$, we recover the GR Lagrangian along with its relevant Newtonian potential.

\begin{figure}[!h]
\centering  \begin{center} \end{center}
\includegraphics[width=0.44 \textwidth,origin=c,angle=0]{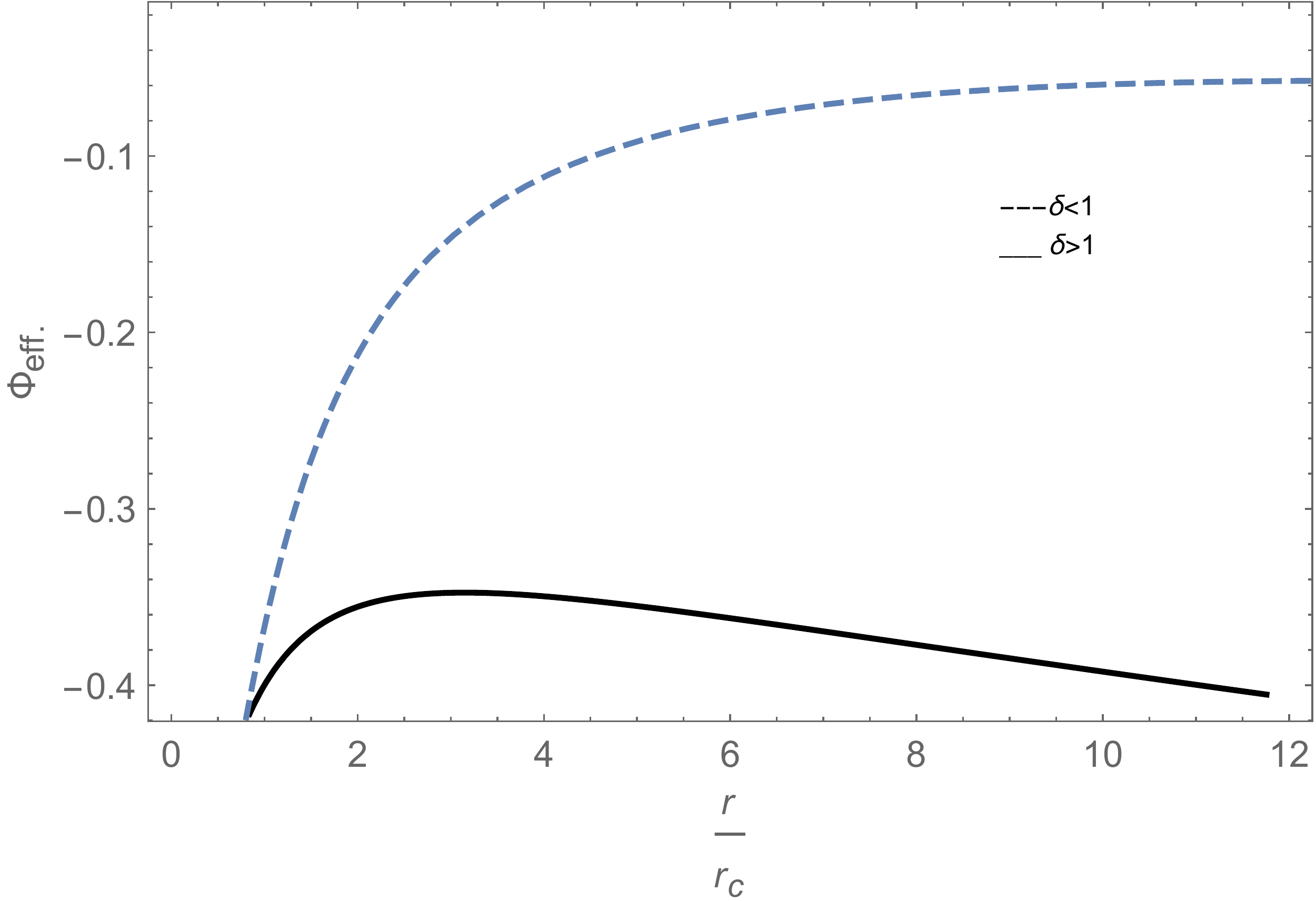}
\caption{\label{fig:p1} The behaviour of $\Phi_{eff}$ diverges for $\delta>1$
(it has an unstable profile) whereas $\delta<1$ has a suitable profile for the interpretation of
galactic dynamics with $\frac{r}{r_c}>1$.}\label{f2}
\end{figure}

The profile of $\delta$ with respect to  $\Phi_{effective}$ can be understood from
Fig. (\textcolor[rgb]{0.00,0.00,1.00}{\ref{f2}}).
Thus, we prefer to work with the small values of model parameter $\delta$.\\

Now, under the conformal transformation $\tilde{r}= \Omega  r $ (obtained through equations
\textcolor[rgb]{0.00,0.00,1.00}{(\ref{a23})} and \textcolor[rgb]{0.00,0.00,1.00}{(\ref{a24})}),
the effective potential (equation \textcolor[rgb]{0.00,0.00,1.00}{({\ref{a27}})})  can be conformally shifted on making use of  equation \textcolor[rgb]{0.00,0.00,1.00}{(\ref{a3})} in the weak field limit and is written as,
\begin{equation}
{\Phi_{effective}(\tilde{r})}\approx -\frac{GM}{2\tilde{r}} \left(1+{\frac{\phi(\tilde{r})}{\sqrt{6} M_{Pl.}}}
\right)\left[ 1+ \left(\frac{\tilde{r}}{\tilde{r}_c}\right)^{\delta}\right].\label{a28}
\end{equation}
Further, withequation \textcolor[rgb]{0.00,0.00,1.00}{(\ref{a3})} under the footnote (1)
for $f(R)=\frac{R^{1+\delta}}{R_{(c)} ^{\delta}}$ with $R_{(c)}$ as a weight constant
having dimension of Ricci Scalar ($R$), we have,
\begin{equation}
R= R_{(c)}\left[\frac{e^{\kappa \phi \sqrt{\frac{2}{3}}}}{1+\delta}\right]^{\frac{1}{\delta}}.\label{a29}
\end{equation}

Now, under the weak field limit, $\kappa \phi<<1$ with $\frac{R}{R_{(c)}}$ normalized to unity, we have
\begin{equation}
\delta= {\frac{2\phi(\tilde{r})}{\sqrt{6} M_{pl.}}}.\label{a30}
\end{equation}
Thus, equation \textcolor[rgb]{0.00,0.00,1.00}{(\ref{a28})} can be rewritten on
using equation \textcolor[rgb]{0.00,0.00,1.00}{(\ref{a30})} as,
 \begin{equation}
{\Phi_{effective}(\tilde{r})}\approx -\frac{GM}{2\tilde{r}} \left(1+\frac{\delta}{2}\right)
\left[ 1+ \left(\frac{\tilde{r}}{\tilde{r}_c}\right)^{\delta}\right]. \label{a31}
\end{equation}\\

The light deflection angle for a spherically symmetric lens (a typical massive spiral galaxy)  with the massive point-like effective potential (equation \textcolor[rgb]{0.00,0.00,1.00}{({\ref{a27}})}), in $f(R)$ background is given as \textcolor[rgb]{0.00,0.00,1.00}{\cite{b38}},
\begin{equation}
{\alpha}= \frac{2GM}{c^2 r_c}\left(\frac{\xi}{r_c}\right)^{-1}\left[ 1+\left(\frac{\xi}{r_c}\right)^{\delta}
\frac{\sqrt{\pi}(1-\delta)\Gamma(1-\frac{\delta}{2})}{2\Gamma(\frac{3}{2}-\frac{\delta}{2})}\right]
,\label{a32}
\end{equation}
where, $\xi$ is a two-dimensional impact parameter, $G$ is the Newtonian gravitational constant
($= 4.3\times 10^{-6}$ kpc km$^{2}$ sec$^{-2}$ $M$ $_{\odot}^{-1}$), $r_c$ is fundamental scaling parameter
in $f(R)$ theory apart from the Schwarzschild length scale \textcolor[rgb]{0.00,0.00,1.00}{\cite{b45}}
and $M$ is galaxy mass in solar mass unit.
The famous classical result of light deflection angle is recovered for $\delta=0$.

For the conformally transformed effective lens potential (equation \textcolor[rgb]{0.00,0.00,1.00}{({\ref{a31}})}), the light deflection angle will be,
\begin{equation}
{\tilde{\alpha}}= \frac{2GM (1+\frac{\delta}{2})}{c^2 \tilde{r}_c}\left(\frac{\tilde{\xi}}
{\tilde{r}_c}\right)^{-1}\left[ 1+\left(\frac{\tilde{\xi}}{\tilde{r}_c}\right)^{\delta}
\frac{\sqrt{\pi}(1-\delta)\Gamma(1-\frac{\delta}{2})}{2\Gamma(\frac{3}{2}-\frac{\delta}{2})}\right].\label{a33}
\end{equation}\\

\begin{figure}[!h]
\centering  \begin{center} \end{center}
\includegraphics[width=0.44 \textwidth,origin=c,angle=0]{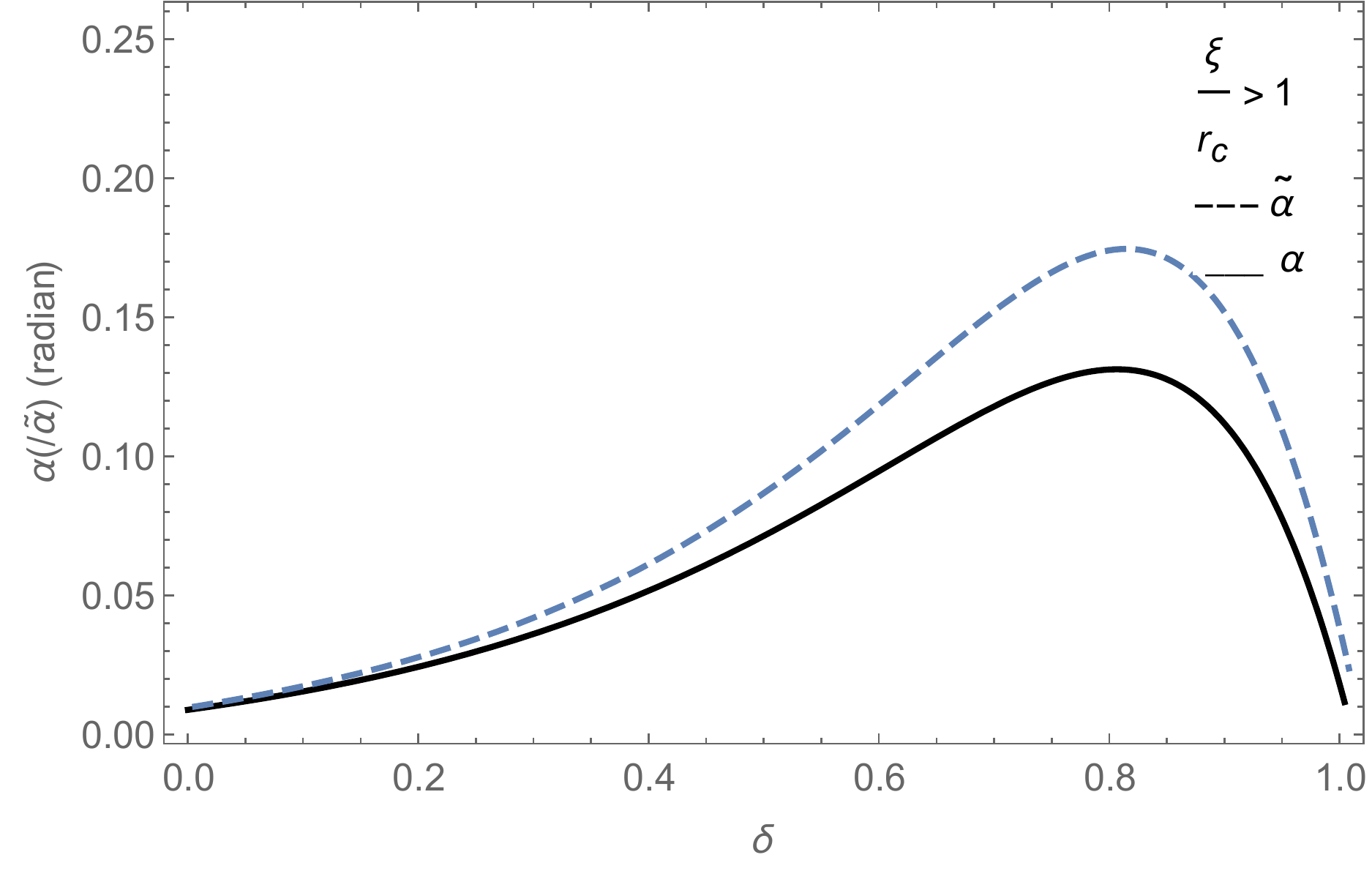}
\caption{\label{fig:p1} The plot shows the behavior of the two conformally related deflection angle profiles w.r.t $\delta<1$ when $\frac{2GM}{c^2 r_c}$ is normalized to unity for the scaled impact parameter, $\frac{{\xi}}{{r}_c}(>1)$.
The solid (Black) curve corresponds to $\alpha$, whereas the dashed curve
corresponds to $\tilde{\alpha}$.  Here, the left wing of both deflection angle curves ($\alpha$ and $\tilde{\alpha}$) seems to be conformally consistent for $\delta<<1$.}\label{f3}
\end{figure}

\begin{figure}[!h]
\centering  \begin{center} \end{center}
\includegraphics[width=0.44 \textwidth,origin=c,angle=0]{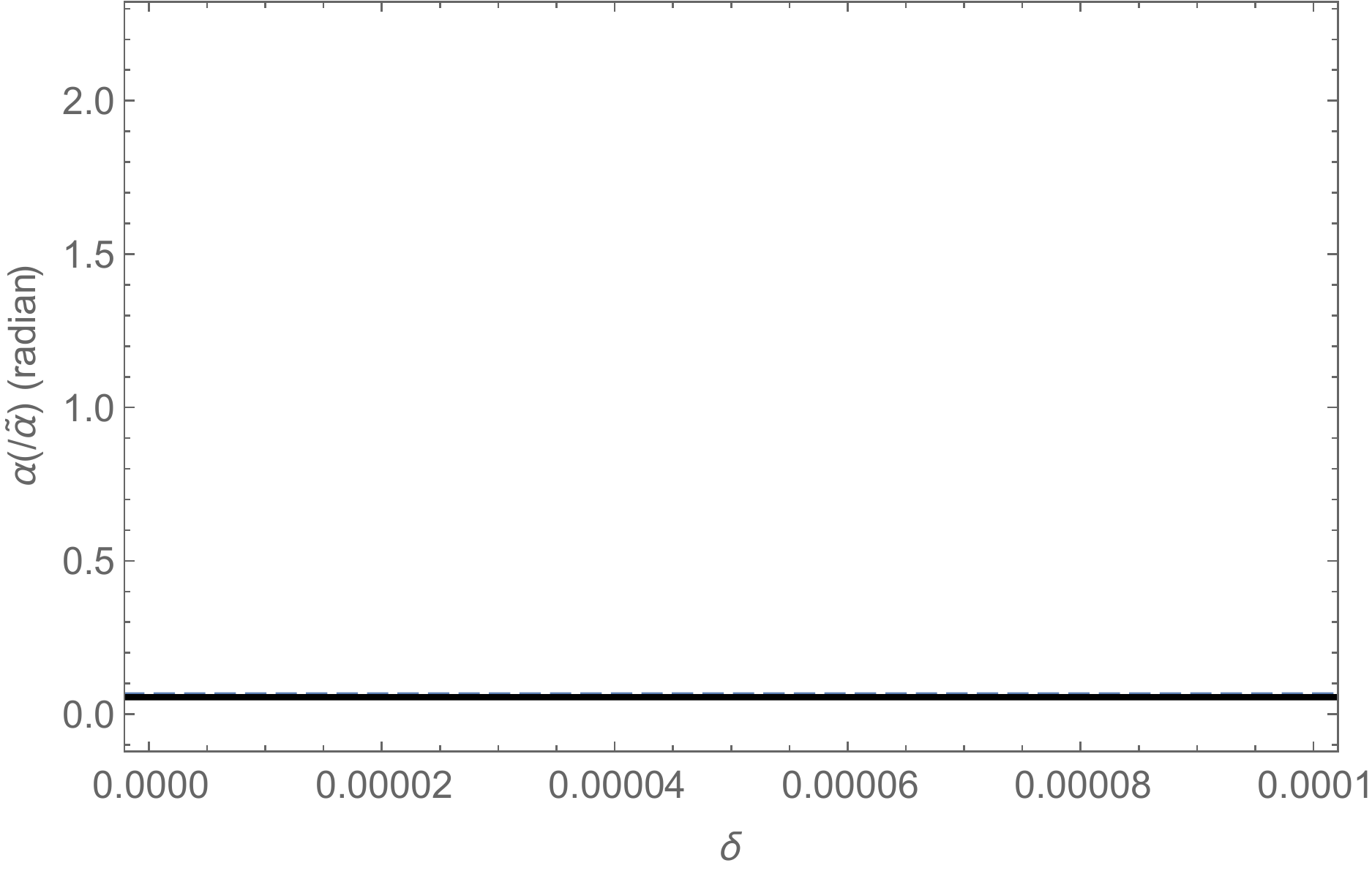}
\caption{\label{fig:p1} The curve is traced for smaller values of $\delta(<<1)$ under the convention used in Fig. 3. It is interpreted from the curve that the profile of
deflection angle seems to be observationally as well as conformally consistent i.e.,
the profile of two deflection angles ($\alpha$ and $\tilde{\alpha}$) are conformally equivalent.}\label{f4}
\end{figure}

The profile of net deflection angles according to the equations \textcolor[rgb]{0.00,0.00,1.00}{(\ref{a32})} and \textcolor[rgb]{0.00,0.00,1.00}{(\ref{a33})} is plotted in
Fig. \textcolor[rgb]{0.00,0.00,1.00}{\ref{f3}} and  Fig. \textcolor[rgb]{0.00,0.00,1.00}{\ref{f4}}.

Thus, it becomes clear by the profile of net light deflection angles  from different figures (\textcolor[rgb]{0.00,0.00,1.00}{\ref{f3}} and \textcolor[rgb]{0.00,0.00,1.00}{\ref{f4}}) to choose the narrow values of $\delta$($<<1$) for conformal equivalence.
The small (about $10^{-6}$) profile of ${\delta}$ is also recently confirmed from our study of galactic dynamics \textcolor[rgb]{0.00,0.00,1.00}{\cite{b33,b34}}.\\
Therefore, for the physical interpretation, we  explore the small values of $\delta$ for the light deflection angles due to the typical massive spiral galaxy system acting as a point-like lens and trace it via the known physical observations of lensing data  for few galaxies culled from \textcolor[rgb]{0.00,0.00,1.00}{\cite{b69}} and references therein.

For instance, the quantitative analysis of the net modified deflection angle can be explored
by considering the case of an  object having solar mass unit (like a typical massive spiral galaxy) in the galactic halo (halo of scalaron cloud due to the $f(R)$ background) acting as a weak lens for the light rays coming from the source star positioned in the external galaxy, like the Magellanic clouds.
\begin{figure}[!h]
\centering  \begin{center} \end{center}
\includegraphics[width=0.44 \textwidth,origin=c,angle=0]{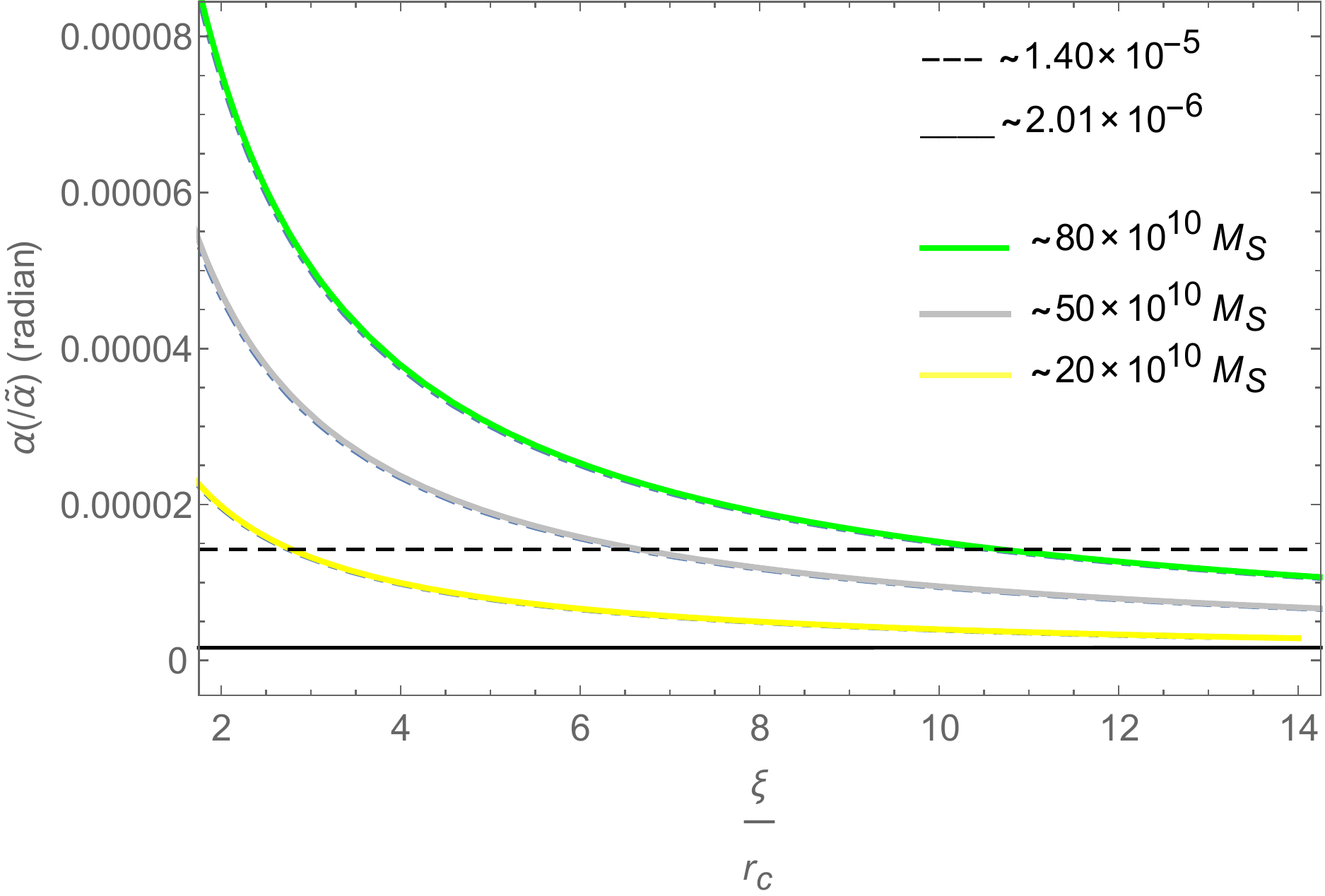}
\caption{\label{fig:p1} The data for typical massive galaxies (as a point-like lenses) is culled from \textcolor[rgb]{0.00,0.00,1.00}{\cite{b69}} and references therein. The plot shows the profiles of net light deflection angles
($\alpha$ (solid colored curves) and $\tilde{\alpha}$ (blue dashed curves imposed over different solid
colored curves)) without any dark matter for different typical massive galaxies with $\delta \approx \mathcal{O} (10^{-6})$ w.r.t the scaled impact parameter.
The deflection angle decreases in the halo of scalaron cloud background (or $f(R)$ background) surrounding the galaxies with increasing value of the scaled impact parameter for different galaxies (see recent works \textcolor[rgb]{0.00,0.00,1.00}{\cite{b34}} and \textcolor[rgb]{0.00,0.00,1.00}{\cite{b69}}).
The horizontal (dashed-black and solid-black) lines represent the observed range of light deflection
angle for different typical massive galaxies.}\label{f5}
\end{figure}
\\Fig. \textcolor[rgb]{0.00,0.00,1.00}{\ref{f5}} shows the profile of deflection angles plotted with smaller value of $\delta$ for different typical massive galaxies acting as a point source.
The observed profile of deflection angle for the galaxies \textcolor[rgb]{0.00,0.00,1.00}{\cite{b69}} is in agreement with the explored $f(R)$ model without dark matter, so the conformally related deflection angles seem to be physically equivalent for $\delta \approx \mathcal{O} (10^{-6})$ \textcolor[rgb]{0.00,0.00,1.00}{\cite{b34,b69}}.

\begin{figure}[!h]
\centering  \begin{center} \end{center}
\includegraphics[width=0.44 \textwidth,origin=c,angle=0]{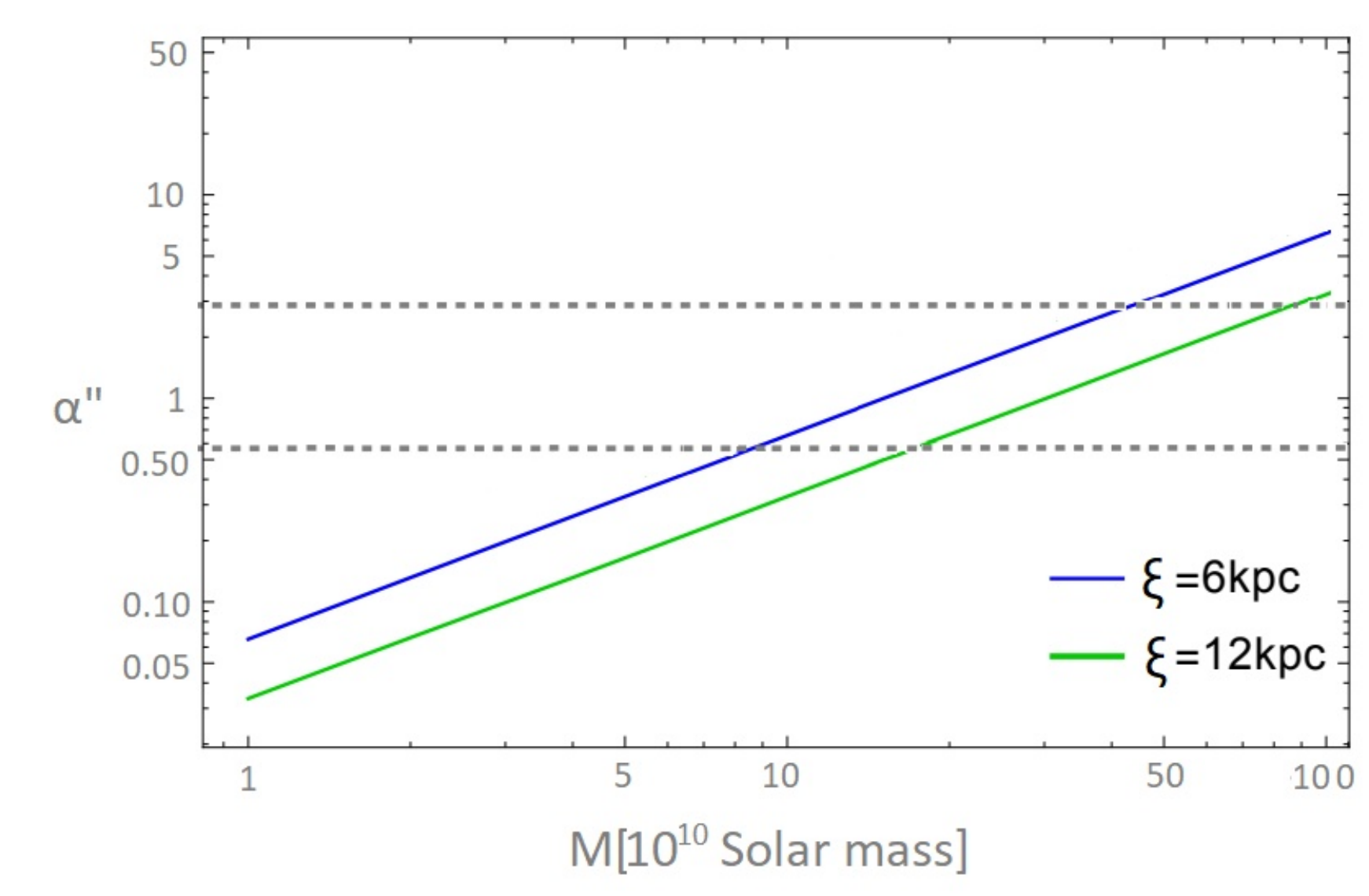}
\caption{\label{fig:p1} The plot shows the profiles of light deflection angles
($\alpha$ in arcsec) for different typical massive galaxies in GR i.e. with $\delta =0$ for different impact parameter ($\xi$).
The deflection angle decreases with increasing value of the impact parameter for different galaxies.
The horizontal (dashed) lines represents the observed range of light deflection
angle for different typical massive galaxies (see recent works \textcolor[rgb]{0.00,0.00,1.00}{\cite{b69}}).}\label{f6}
\end{figure}

We have also plotted in Fig.\textcolor[rgb]{0.00,0.00,1.00}{\ref{f6}}  the deflection angles for typical massive galaxies in GR.

As an important point to draw attention here is that  the extent of $\delta$ can also be probed directly via the study of orbital motion of neutral hydrogen clouds (as a test particle) in stable orbits far from the visible profile of typical spiral galaxies with $\delta=v_{orbital}^2/c^2$ taking $\mid v_{orbital} \mid \approx 220-300$ km/s \textcolor[rgb]{0.00,0.00,1.00}{\cite{b36}}.

\section{\label{4}$f(R)$ conformal non-null geodesics analysis and phenomenology for perihelion shift}

Because of the uncertainty in the GR predicted result of Mercury's orbit \textcolor[rgb]{0.00,0.00,1.00}{\cite{b62,b63,b64,b65}} as well as the upcoming probes \textcolor[rgb]{0.00,0.00,1.00}{\cite{b66}},  motivate us to study the Mercury's orbit in $f(R)$ theory in order to investigate the bare effect of deviation ($\delta$) at the planetary scale, i.e. $R\rightarrow R^{1+\delta}$.

Earlier investigation for the present case was done with Yukawa-like potential in $f(R)$ theory \textcolor[rgb]{0.00,0.00,1.00}{\cite{b38}}. Unlike deflection of light, here the case is that an object (test mass) never gets out to infinity where the spacetime metric is asymptotically Minkowskian. Hence, for the study of $f(R)$ conformal non-null geodesics, we start from equation \textcolor[rgb]{0.00,0.00,1.00}{(\ref{a18})}. Under this case, the last term of equation
\textcolor[rgb]{0.00,0.00,1.00}{(\ref{a18})} does not vanish for the non-relativistic
material particles having non-zero rest mass and so it is not further possible to reduce the equation
\textcolor[rgb]{0.00,0.00,1.00}{(\ref{a18})} to the original one \textcolor[rgb]{0.00,0.00,1.00}{(\ref{a15})}
by different conformal relations like equation \textcolor[rgb]{0.00,0.00,1.00}{(\ref{a21})}.
Thus, equation \textcolor[rgb]{0.00,0.00,1.00}{(\ref{a18})} suggests that under the conformally related
$f(R)$ spacetime (equation \textcolor[rgb]{0.00,0.00,1.00}{(\ref{a3})}),
the non-null trajectories experiences an extra force for unit test mass. This extra force can be the carrier of a new particle. In $f(R)$ gravity theory, this new particle will be the scalar field particle called scalaron (see reference \textcolor[rgb]{0.00,0.00,1.00}{\cite{b72}}).

Let us investigate the non-relativistic case of equation \textcolor[rgb]{0.00,0.00,1.00}{(\ref{a18})}
for the material particle (non-zero rest mass) tracing the curve $x^\gamma (\tau)$. For mathematical convenience,
we normalize its last term,
$g_{\mu\nu}\frac{d{x^\mu}}{d\tau}\frac{d{x^\nu}}{d\tau}=-1$ \textcolor[rgb]{0.00,0.00,1.00}{\cite{b73}}.
Thus, equation \textcolor[rgb]{0.00,0.00,1.00}{(\ref{a18})} can be re-written as,
\begin{eqnarray}
\begin{split}
\frac{d^2{x^\gamma}}{d\tau^2}+\Gamma^{\gamma}_{\mu \nu}\frac{d{x^\mu}}{d\tau}\frac{d{x^\nu}}{d\tau}+
2\frac{d}{d\tau}(\ln \Omega) \frac{d{x^\gamma}}{d\tau}\\
+\partial^\gamma (\ln \Omega)=0
,\label{a34}
\end{split}
\end{eqnarray}

In the non-relativistic limit with $\gamma=i$ (space coordinates) one should get the
Newtonian equation with the conventional gravitational force \textcolor[rgb]{0.00,0.00,1.00}{\cite{b73}}.
Therefore, equation
\textcolor[rgb]{0.00,0.00,1.00}{({\ref{a34}})} has the following form,
\begin{eqnarray}
\begin{split}
\frac{d^2{x^i}}{d\tau^2}+\Gamma^{i}_{\mu \nu}\frac{d{x^\mu}}{d\tau}\frac{d{x^\nu}}{d\tau}+
2\frac{d}{d\tau}(\ln \Omega) \frac{d{x^i}}{d\tau}\\
+\partial^i (\ln \Omega)=0
,\label{a35}
\end{split}
\end{eqnarray}

So, under the non-relativistic Newtonian limit,
equation \textcolor[rgb]{0.00,0.00,1.00}{(\ref{a35})} becomes,

\begin{eqnarray}
\frac{d^2{x^i}}{d\tau^2}=-[\partial^i \Phi + \partial^i (\ln \Omega)]=
-[\vec{\nabla}_{3D}\Phi + \vec{\nabla}_{3D} (\ln \Omega) ].\label{a36}
\end{eqnarray}

Now, if $-\vec{\nabla}_{3D}\Phi$ is the conventional (or standard) force of gravity, then under the
conformal transformation of spacetime metric (equation \textcolor[rgb]{0.00,0.00,1.00}{(\ref{a3})}),
we have an additional force on the unit test mass,
\begin{eqnarray}
{\vec{F}_{\Omega}}=-\vec{\nabla}_{3D} (\ln \Omega),\label{a37}
\end{eqnarray}
where $\Omega=\left(\frac{\partial f(R)}{\partial R}\right)^{\frac{1}{2}}$ is the general conformal factor of the $f(R)$ theory.

Clearly, the extent of such extra force will depend upon the extent of $f(R)$ model parameter ($\delta$) and vanishes for $\delta=0$. Also, in contrast to the Newtonian gravitational force, it die off in strength slowly as,

\begin{eqnarray}
{\vec{F}_{\Omega}}=-\frac{f_{RR}}{2f_R} R_r=\frac{\delta(3-\delta)}{2r}\hat{r},\label{aa37}
\end{eqnarray}
where, $f_{RR}=\frac{\partial F}{\partial R}$, $R_r=\frac{\partial R}{\partial r}$ and $R$ is the Ricci scalar curvature corresponding to the metric equations \textcolor[rgb]{0.00,0.00,1.00}{(\ref{a23})} with \textcolor[rgb]{0.00,0.00,1.00}{(\ref{a27})}. It becomes clear from equation \textcolor[rgb]{0.00,0.00,1.00}{(\ref{aa37})} that the suppression of such extra force depends on the constrained values of $\delta$ which from the observations in high-density regions is usually much small.

It is concluded from the non-vanishing value of $R$ given as,
\begin{eqnarray}
R=\delta (3-\delta) \frac{GM}{r^3},\label{aaa37}
\end{eqnarray}
that outside the source,  value of Ricci curvature scalar $R$ depends on the model parameter $\delta$ and vanish for $\delta=0$ in GR theory and hence the corresponding force also vanish. As it is well known that a deviation from the $\frac{1}{r^2}$ Newtonian force is responsible to a perihelion advancement of Mercury's orbit (explained observationally by Einstein's gravity theory). Therefore, any modification in the Einstein's GR theory must be much smaller for interpreting phenomenologically the present case.

Now, for the phenomenological study of $f(R)$ conformal non-null geodesics under an idealized consideration of system, we explore the time-like orbit of Mercury at the Solar system scale.

We simplify the problem in $f(R)$ theory by using the corresponding symmetries and write the modified gravity Lagrangian following equation \textcolor[rgb]{0.00,0.00,1.00}{(\ref{a23})} as,
\begin{equation}
2L= -(1+2\Phi_{effective}) \dot{t}^{2} + (1+2\Phi_{effective})^{-1} \dot{r}^{2} + r^2 \dot{\varphi}^{2},\label{a38}
\end{equation}
where $\Phi_{effective}$ is given by equation \textcolor[rgb]{0.00,0.00,1.00}{(\ref{a25})} and the dot denotes
differentiation w.r.t the proper time ($\tau$).

The $(\varphi)$  and $(t)$  equations following from this Lagrangian are given by using Euler's Lagrangian
formalism as,
\begin{equation}
 r^2 \dot{\varphi}=l,\label{a39}
\end{equation}
and
\begin{equation}
(1+2\Phi_{effective}) \dot{t}=k,\label{a40}
\end{equation}
where, $l$ and $k$ are constants.
The $(r)$ equation can be simplified for material particle with $c=1$ as,
\begin{equation}
1= -(1+2\Phi_{effective}) \dot{t}^{2} + (1+2\Phi_{effective})^{-1} \dot{r}^{2} + r^2 \dot{\varphi}^{2}.\label{a41}
\end{equation}

Dividing throughout by $\dot{\varphi}^{2}$, and using  equations \textcolor[rgb]{0.00,0.00,1.00}{(\ref{a39})}
and \textcolor[rgb]{0.00,0.00,1.00}{(\ref{a40})}, we get,

\begin{equation}
\left(\frac{dr}{d\varphi}\right)^2= {-r}^2 (1+2\Phi_{effective}) +
\frac{r^4}{l^{2}} \left( k^{2}+1+ 2\Phi_{effective}\right).\label{a42}
\end{equation}

Now, on defining a new variable as,
\begin{equation}
u\equiv\frac{1}{r},\label{a43}
\end{equation}
and
\begin{equation}
\left(\frac{dr}{d\varphi}\right)= -\frac{1}{u^2} \frac{du}{d\varphi}.\label{a44}
\end{equation}

Thus, equation \textcolor[rgb]{0.00,0.00,1.00}{(\ref{a42})} can be written as,
\begin{equation}
\begin{split}
\left(\frac{du}{d\varphi}\right)^2= \frac{1}{l^2}\left[ (k^2 + 1)-2GMu\left( \frac{1}{2}+\frac{1}{2}
\frac{u^{-\delta}}{r_c^{\delta}}\right)\right]-\\u^2\left[1-2 GMu\left(\frac{1}{2}+\frac{1}{2}
\frac{u^{-\delta}}{r_c^{\delta}}\right)\right],\label{a45}
\end{split}
\end{equation}
where $M$ is the mass of the Sun.

In contrast to the Newtonian classical equation of motion \textcolor[rgb]{0.00,0.00,1.00}{\cite{b73}}, we have
\begin{equation}
\left(\frac{du}{d\varphi}\right)^2= \frac{2E}{l^2}+ \frac{2GM}{l^2}u-u^2,\label{a46}
\end{equation}

we can rewrite the modified relativistic equation \textcolor[rgb]{0.00,0.00,1.00}{(\ref{a45})} as,
\begin{equation}
\begin{split}
\left(\frac{du}{d\varphi}\right)^2= \frac{2E}{l^2}+ \left(\frac{-2GM}{l^2}\right) u \left(\frac{1}{2}+\frac{1}{2}
\frac{u^{-\delta}}{r_c^{\delta}}\right)-u^2 + \\2GMu^3\left(\frac{1}{2}+\frac{1}{2}
\frac{u^{-\delta}}{r_c^{\delta}}\right),\label{a47}
\end{split}
\end{equation}
where, $2E\equiv k^2 +1$ with $E$ as total energy of the system. We observe that the relativistic correction factor (the last term which is responsible for the advancement of perihelion motion of a planet) is also enhanced by modification. Therefore, the $\delta$ deviated (i.e., $R \rightarrow R^{1+\delta}$) $f(R)$ modifications must be smaller in order not to violate the GR result but possible probe the uncertainty found if any by different current running projects and to have an agreement with the observation. Such modification may be compatible for explaining an uncertainty in the GR prediction under the current (MESSENGER)\textcolor[rgb]{0.00,0.00,1.00}{\cite{b65}} and planned (the European-Japanese BepiColombo) \textcolor[rgb]{0.00,0.00,1.00}{\cite{b66}} missions.\\

If we express $2GM$ with $\epsilon$ as a Schwarzschild length in units of $c=1$ and
$\bar{{\epsilon}} =\epsilon \left(\frac{1}{2}+\frac{1}{2}\frac{u^{-\delta}}{r_c^{\delta}}\right)$ as a modified
Schwarzschild length, then \textcolor[rgb]{0.00,0.00,1.00}{(\ref{a47})} can be written as,
\begin{equation}
\left(\frac{du}{d\varphi}\right)^2= \bar{{\epsilon}} u^3-u^2-\bar{{\epsilon}} \frac{u}{l^2} + \frac{2E}{l^2}.
\label{a48}
\end{equation}

Clearly, for $\delta=0$, we recover the GR differential equation of motion and hence, the shift in perihelion
can be obtained via the standard perturbation approach to the Newtonian solution \textcolor[rgb]{0.00,0.00,1.00}{\cite{b73}}.\\

Now, to determine the shift in perihelion, we  solve equation \textcolor[rgb]{0.00,0.00,1.00}{(\ref{a48})} by equating it to zero, since for perihelion and aphelion $\varphi$ is fixed. Actually, this will suggest us different positions.
Since equation \textcolor[rgb]{0.00,0.00,1.00}{(\ref{a48})} is a cubic equation, so has $u_1, u_2$ and $u_3$ as its three different solutions or roots. But the semi-classical observed problem has bounded solution between perihelion $(u_1)$ and aphelion $(u_2)$. Thus, among the three solutions, one solution must be an unphysical.
Therefore, if we assume that $u_3$ as an unphysical solution, then we must eliminate it from
equation \textcolor[rgb]{0.00,0.00,1.00}{(\ref{a48})} for the study of bounded motion.

From equation \textcolor[rgb]{0.00,0.00,1.00}{(\ref{a48})}, it is possible to replace the R.H.S. with the three roots viz., $u_1$, $u_2$ and $u_3$ as,
\begin{equation}
\left(\frac{du}{d\varphi}\right)= [\bar{{\epsilon}}(u-u_1)(u-u_2)(u-u_3)]^{\frac{1}{2}}.\label{a49}
\end{equation}

Under the assumption of bounded motion, $u_1\leq u \leq u_2$, equation
\textcolor[rgb]{0.00,0.00,1.00}{(\ref{a49})} can be rewritten as,
\begin{equation}
\left(\frac{du}{d\varphi}\right)= [\bar{{\epsilon}}(u-u_1)(u_2-u)(u_3 -u)]^{\frac{1}{2}},\label{a50}
\end{equation}

To eliminate the assumed unphysical solution i.e., $u_3$ from equation \textcolor[rgb]{0.00,0.00,1.00}{(\ref{a50})},
we must impose a boundary condition on the three solutions.

Since $u$ has the dimension of inverse of length and $\bar{\epsilon}$ has the dimension of length. Because $u_1$ and $u_2$ are basically considered as two roots of equation \textcolor[rgb]{0.00,0.00,1.00}{(\ref{a46})}, so $u_1+u_2$ represents the inverse length of major axis \textcolor[rgb]{0.00,0.00,1.00}{\cite{b73}} and if $u_3$ is positive then $u_1+u_2+u_3$ must be a large in contrast to modified Schwarzschild length $\bar{{\epsilon}}$. So, the plausible boundary condition can be constructed as,
\begin{equation}
u_1 + u_2 + u_3 =\frac{1}{\bar{{\epsilon}}}.\label{a51}
\end{equation}

From equations \textcolor[rgb]{0.00,0.00,1.00}{(\ref{a50})} and \textcolor[rgb]{0.00,0.00,1.00}{(\ref{a51})}, for
the bounded motion, the change in the angle between $u_1$ and $u_2$ is,
\begin{eqnarray}
\begin{split}
\mid\triangle\varphi\mid\approx\int_{u_1}^{u_2}
{\left(\frac{1}{(u-u_1)(u_2-u)}\right)^{\frac{1}{2}}\left(1+\frac{\bar{\epsilon}}{2}(u+u_1+u_2)\right)\ du}
\label{a52}
\end{split}
\end{eqnarray}

Let us now simplify equation \textcolor[rgb]{0.00,0.00,1.00}{(\ref{a52})} by defining two parameters as,
\begin{equation}
\alpha \equiv\frac{1}{2}[u_1+u_2],\label{a53}
\end{equation}
and
\begin{equation}
\beta \equiv\frac{1}{2}[u_2-u_1],\label{a54}
\end{equation}

Thus following equations \textcolor[rgb]{0.00,0.00,1.00}{(\ref{a53})} and
\textcolor[rgb]{0.00,0.00,1.00}{(\ref{a54})}, we have $u_1=\alpha-\beta$, $u_2=\alpha+\beta$ and
$(u-u_1)(u_2-u)=[\beta^2-(u-\alpha)^2]$.\\

Now, equation \textcolor[rgb]{0.00,0.00,1.00}{(\ref{a52})} become,
\begin{eqnarray}
\mid\triangle \varphi\mid\approx\int_{u_1}^{u_2}
{\left(\frac{1+\frac{\bar{\epsilon}}{2}u + \bar{\epsilon} \alpha}{[\beta^2-(u-\alpha)^2]^\frac{1}{2}}\right)\ du}
\label{a55}.\end{eqnarray}

Further, equation \textcolor[rgb]{0.00,0.00,1.00}{(\ref{a55})} can be solved to give,

\begin{eqnarray}
\mid\triangle \varphi\mid\approx
\left[-\frac{1}{2}\bar{\epsilon}(\beta^2-(u-\alpha)^2)^{\frac{1}{2}}+
\left(1+ \frac{3\bar{\epsilon}}{2}\alpha\right) {\sin}^{-1}\frac{u-\alpha}{\beta}\right]_{u_1}^{u_2}
\label{a56}.\end{eqnarray}

By using equations \textcolor[rgb]{0.00,0.00,1.00}{(\ref{a53})} and \textcolor[rgb]{0.00,0.00,1.00}{(\ref{a54})},
we get from equation \textcolor[rgb]{0.00,0.00,1.00}{(\ref{a56})},
\begin{eqnarray}
\mid\triangle \varphi\mid\approx \pi \left( 1+ \frac{3}{2}\bar{\epsilon}\alpha\right)
\label{a57}.\end{eqnarray}

The twice of the left hand side of the equation \textcolor[rgb]{0.00,0.00,1.00}{(\ref{a57})} suggests the angle between successive
perihelion. The perihelion shift can be computed as,
 \begin{eqnarray}
\Delta\varphi=2\mid\triangle \varphi\mid-2\pi
\label{a58}.\end{eqnarray}\\

Thus, the required perihelion shift in $f(R)$ gravity theory is,
\begin{eqnarray}
\begin{split}
\Delta\varphi=3\pi\bar{\epsilon}\alpha=3\pi(2GM)\left(\frac{1}{2}+\frac{1}{2}{\left(\frac{r}{r_c}\right)}^{\delta}
\right)\times
\\ \left(\frac{1}{2}(u_1+u_2)\right)
\label{a59}.
\end{split}
\end{eqnarray}

The last factor of above equation can be classically given from the values of perihelion and aphelion in terms of semi-major axis and eccentricity \textcolor[rgb]{0.00,0.00,1.00}{\cite{b73}}. So, we can rewrite equation \textcolor[rgb]{0.00,0.00,1.00}{(\ref{a59})} under the observed bounded motion as,

\begin{eqnarray}
\Delta\varphi=\frac{6\pi G M \left(\frac{1}{2}+\frac{1}{2}{\left(\frac{r}{r_c}\right)}^{\delta}\right)}{c^2 a(1-e^2)}
\label{a60},
\end{eqnarray}
where $a$ is the semi-major axis, $e$ is the eccentricity of the orbit which for a planet like Mercury are
available from the observations and $c$ is the speed of light. The last factor in equation \textcolor[rgb]{0.00,0.00,1.00}{(\ref{a60})} in numerator indicates that deviation $\delta$, which arises due to modification in Einstein gravity under $R \rightarrow R^{1+\delta}$, contributes to the perihelion shift within the permissible limit and therefore must be much smaller. Thus, the effect of such modification may probe the uncertainty in GR result as suggested after different missions \textcolor[rgb]{0.00,0.00,1.00}{\cite{b62,b63,b64,b65,b66}}.

We investigate equation \textcolor[rgb]{0.00,0.00,1.00}{(\ref{a60})} for $\delta=0$ with the values of
$a \approx 58\times 10^9$ m  and $e \approx 0.2056$ for Mercury precession about the massive Sun,
$M \approx 1.989\times10^{30}$ kg and $G = 6.672\times10^{-11}$ m$^{3}$/(kg-sec$^{-2}$). The data is culled from NASA's Solar System Bodies \textcolor[rgb]{0.00,0.00,1.00}{\cite{b74}}.
With these values we get, $\Delta\varphi \approx 0.5 \times 10^{-6}$ radian per revolution or
$\approx 0.10312^{\prime\prime}$ per revolution. As, the periodic time of Mercury around the Sun
is $88$ Earth-days so that it makes $\frac{365}{88} \approx 4.148$ revolution per year,
or $415$ revolutions in one century.  Therefore, the advance of the perihelion of Mercury in 1 century is
$0.10312\times 415 \approx 42.79$ seconds of arc.
Thus, the GR result is fully recovered.

The relativistic conformal shift of the perihelion of Mercury orbit can be written
(following equation \textcolor[rgb]{0.00,0.00,1.00}{(\ref{a29})} and \textcolor[rgb]{0.00,0.00,1.00}{(\ref{a30})})
as,
\begin{eqnarray}
\Delta\tilde{\varphi}=\left(1+\frac{\delta}{2}\right)\frac{6\pi G M \left(\frac{1}{2}+
\frac{1}{2}(\frac{\tilde{r}}{\tilde{r}_c})^{\delta}\right)}{c^2 a(1-e^2)}
\label{a61}.
\end{eqnarray}\\

The profile of perihelion shift according to the equations \textcolor[rgb]{0.00,0.00,1.00}{(\ref{a60})}
and \textcolor[rgb]{0.00,0.00,1.00}{(\ref{a61})} is plotted in Fig. \textcolor[rgb]{0.00,0.00,1.00}{7} and
Fig. \textcolor[rgb]{0.00,0.00,1.00}{8} for different values of $f(R)$ model parameter $\delta$.
\begin{figure}[!h]
\centering  \begin{center} \end{center}
\includegraphics[width=0.44 \textwidth,origin=c,angle=0]{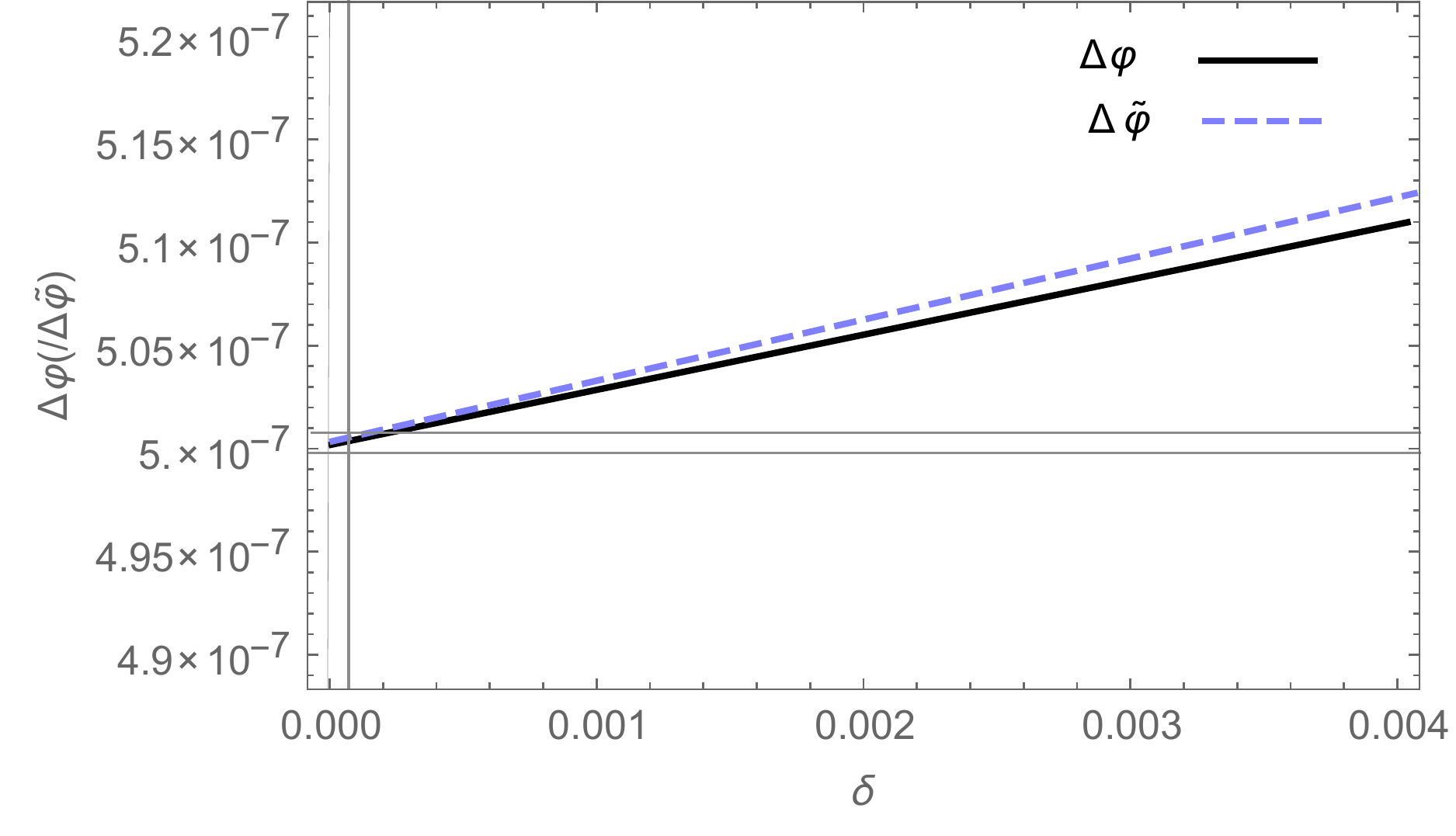}
\caption{\label{fig:p8} The plot shows the behaviour of perihelion shifts for $\delta<1$ with $\frac{r}{{r}_c}>1$.
The solid (Black) curve shows that $\Delta{\varphi}$ increases with increasing values of $\delta$ whereas
$\Delta\tilde{\varphi}$ (the dashed curve) increases rapidly in contrast to the  solid (Black) curve
w.r.t smaller $\delta$. The narrow vertical strip corresponds to the allowed smaller value of $\delta$ for conformal equivalence whereas the horizontal narrow strip corresponds to the observed value \textcolor[rgb]{0.00,0.00,1.00}{\cite{b74,b62,b63,b64,b65}}.}\label{f7}
\end{figure}

It becomes clear from  Fig. \textcolor[rgb]{0.00,0.00,1.00}{7} that the perihelion shifts ($\Delta{\varphi}$ and $\Delta\tilde{\varphi}$) vary largely  w.r.t the small deviation in the $f(R)$ model parameter and also not observationally consistent with $\delta<1$ .
But, from Fig. \textcolor[rgb]{0.00,0.00,1.00}{8}, it is clear that for $\delta<<1$ the perihelion shift approximately attains an observable constant value  for $\Delta{\varphi}$ and $\Delta\tilde{\varphi}$.
\begin{figure}[!h]
\centering  \begin{center} \end{center}
\includegraphics[width=0.44 \textwidth,origin=c,angle=0]{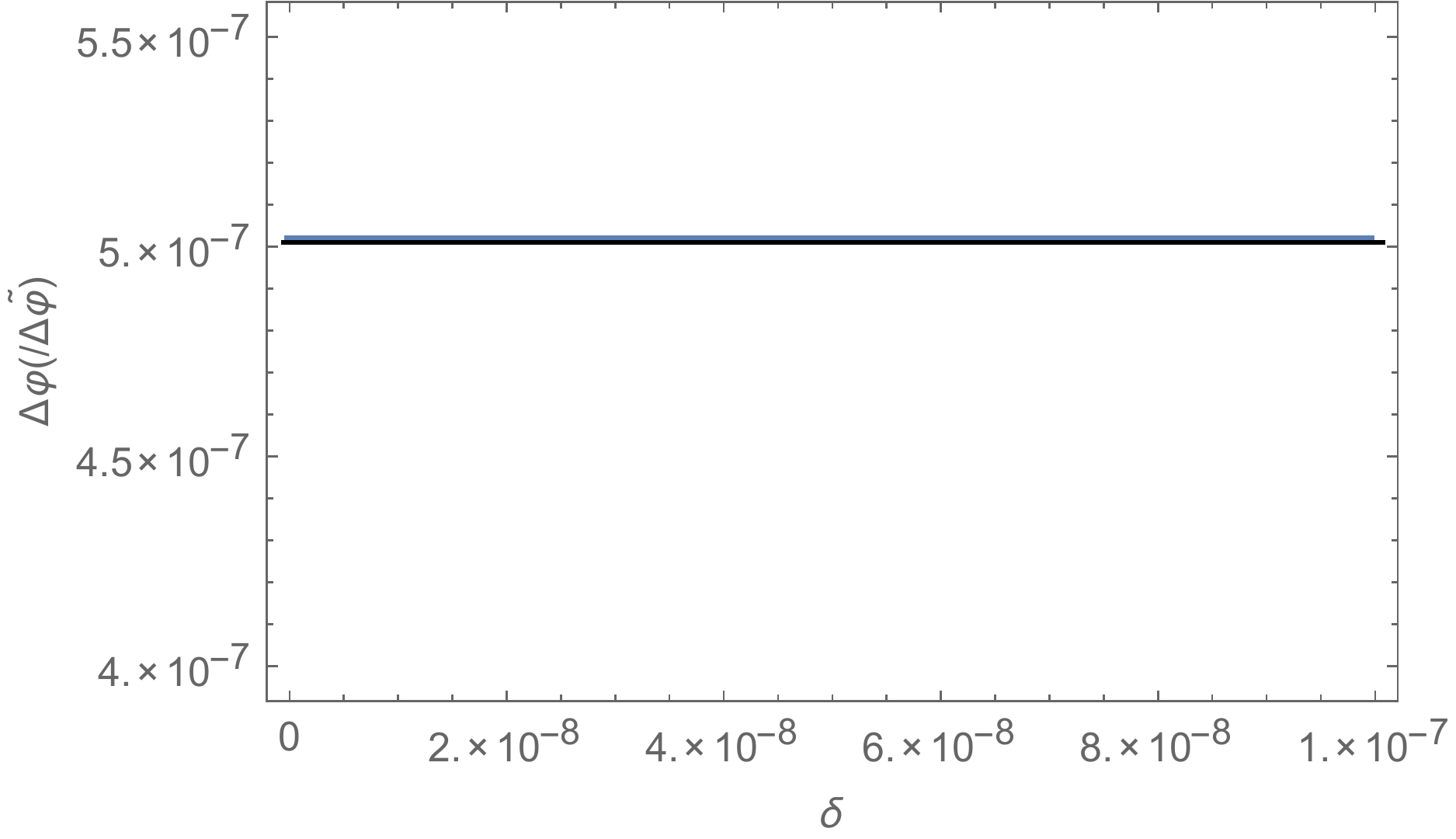}
\caption{\label{fig:p9} It becomes clear from the plot that for $\delta<<1$, the perihelion shifts of Mercury orbit seems to explain the observed result ($42.98^{\prime\prime}$ per century).
}\label{f8}
\end{figure}

As a further interesting fact, we also look for the number of revolutions corresponding to the age of our Solar System. For instance, for the $1$ degree shift of the orbit, the time taken will be $\frac{3600}{42.98}\approx 84$ centuries. Now, for 360 degrees, the required time is $\approx 3.028$ millions of years (or $3,028,037$ years). Obviously, this is not very much at cosmological time scale. In about $5$ billion years (during which Sun, Earth and other planets form) before the present epoch, the number of revolutions the axis of Mercury has gone through is $\frac{5\times 10^9}{3.028\times 10^6}\approx 1651$!

According to Fig. \textcolor[rgb]{0.00,0.00,1.00}{8}, the perihelion shift can be successfully explained for $\delta\approx \mathcal{O} (10^{-7})$. Such constraint can be directly probed via specifying the model parameter $\delta$ with the orbital velocity of Mercury ($48$ km/s) as $\delta=v_{orbital}^2/c^2$. Contrasting with the constraints explored by Zakharov et al., ( $\delta \approx 10^{-13}$) and Clifton ( $\delta \approx 10^{-19}$) \textcolor[rgb]{0.00,0.00,1.00}{\cite{b37, b037}}, the explored constraint is different from them and corresponds to the extra force which drops off as $r^{-1}$.  Such relativistic effects is likely to be detectable by different probes as discussed by Will in reference \textcolor[rgb]{0.00,0.00,1.00}{\cite{b66}}. It in also in close approximation with the parameterized post-Newtonian (PPN) estimates.

Thus, with such profile of $\delta$ (much small value) (Fig. \textcolor[rgb]{0.00,0.00,1.00}{8}), the additional fifth force given by equation \textcolor[rgb]{0.00,0.00,1.00}{(\ref{a37})} can be approximated as a small constant force  at the Solar system scales and can be screened \textcolor[rgb]{0.00,0.00,1.00}{\cite{b27,b75}}.

\section{\label{7} Realization of late-time cosmic constraints via bare $f(R)$ self-interaction potential}

The recent (2018) Planck Collaboration \textcolor[rgb]{0.00,0.00,1.00}{\cite{b1,b2}} has measured the Equation of State parameter (EoS) $\omega=-1.028\pm0.032$, which  motivates us to explore an alternative theory of gravity and also because of the serious issues with the classical cosmological constant $\Lambda$ with $\omega=-1$ \textcolor[rgb]{0.00,0.00,1.00}{\cite{b3}}. Actually, the case of positive cosmological constant $\Lambda$ (also interpreted as vacuum energy density) can be viewed in two different perspectives till 1998 (its smallness) and after 1998 (its coincidence with present average matter density of the universe). At present, the common view that at the background level there is nothing but a cosmological constant of positive nature and an alternative gravity theory may work at large scales because of the observations demanding the EoS not exactly equal to $-1$. We investigate the present $f(R)$ model in this connection and explore the constraint via the vacuum solution of the theory with bare conformal $f(R)$ potential.

To see the extent of power-law $f(R)$ curvature scalar in Jordan frame fitted with the cosmic observational data, Capozziello et al.,  explored the encouraging results with SNIa and WMAP data and constrained the cosmic viable shifted power-law $f(R)$ model parameter, $\delta$ to be approximately -0.6 or 0.4 \textcolor[rgb]{0.00,0.00,1.00}{\cite{b77}}.

Now, to realize the observed late-time cosmic constraint on the shifted $f(R)$ model parameter in the conformal Einstein frame, we argue that it is the form of bare $f(R)$ curvature scalar potential (equation \textcolor[rgb]{0.00,0.00,1.00}{(\ref{a6})}) that provides the current cosmic acceleration.  So, accordingly we can write the observed dark energy parameter (positive Cosmological constant) as \textcolor[rgb]{0.00,0.00,1.00}{\cite{b76,b75,b72,b27}},

\begin{eqnarray}
\Lambda_{obs.}=\displaystyle\left\lvert \kappa^2 V(\phi) \right\rvert=\displaystyle\left\lvert \frac{R f_R-f(R)}{2 (f_R)^2}\right\rvert _{R\rightarrow R_{(c)}}
\label{a62}.
\end{eqnarray}

Here $\Lambda_{obs.}$ is the observed dark energy parameter and $R_c$ is the weight constant of our model which we assign here to be the background mass scale and may represents the  critical density or vacuum energy density in the universe today \textcolor[rgb]{0.00,0.00,1.00}{\cite{b27}}. During the late-times $R$  saturates (or dilutes) to $R_{(c)}$ (cosmic coincidence). Such coincidence can be seen as  problematic only if one starts assuming that one could be present with equal probability in any of the periods of the cosmic evolution. But this is definitely not the case \textcolor[rgb]{0.00,0.00,1.00}{\cite{b3,b27}}.

Before exploring \textcolor[rgb]{0.00,0.00,1.00}{(\ref{a62})} under above discussion, we study the extremum condition on the Einstein frame bare $f(R)$ potential and obtain the bound on the model parameter $\delta$.
Because it is the bare $f(R)$ potential which decides the extremum conditions for the effective $f(R)$ potential (i.e., including non-relativistic matter fields) under a suitably chosen function $f(R)$ in Einstein frame \textcolor[rgb]{0.00,0.00,1.00}{\cite{b35}}. On using equations \textcolor[rgb]{0.00,0.00,1.00}{(\ref{a3})} and \textcolor[rgb]{0.00,0.00,1.00}{(\ref{a6})} and the Footnote $1$, we have

\begin{eqnarray}
\frac{dV(\phi)}{d \phi}=\frac{M_{Pl}}{\sqrt{6}{f_R}^2}[Rf_R-2f]
\label{a63},
\end{eqnarray}

Clearly, for the minima of effective $f(R)$ potential, we must have $\frac{dV(\phi)}{d \phi}<0$ and $\frac{d^2V(\phi)}{d \phi^2}>0$. For $\delta<1$, the extremum conditions are satisfied as
\begin{eqnarray}
\frac{dV(\phi)}{d \phi}= \frac{M_{Pl}}{\sqrt{6}}\frac{(\delta-1)R}{(\delta+1)^2}\left(\frac{R_{(c)}}{R}\right)^{\delta}<0
\label{a630},
\end{eqnarray}

\begin{eqnarray}
 \frac{d^2V(\phi)}{d \phi^2}=\frac{1}{3}\left[ \frac{R}{f_R}+\frac{1}{f_{RR}}-
 \frac{4f}{(f_R)^{2}}\right]>0
\label{a64},
\end{eqnarray}

Therefore, on further simplifying equation \textcolor[rgb]{0.00,0.00,1.00}{(\ref{a62})} with small $\delta$, we get

\begin{eqnarray}
R_{\Lambda_{obs.}} \approx \displaystyle\left\lvert \frac{\delta}{2 (1+2\delta)} R_{(c)}\right\rvert
\label{a65}.
\end{eqnarray}

Clearly, for $\delta=0$, the $f(R)$ contribution vanishes and now we have to add by hand a classical cosmological constant in GR geometric action for interpreting present cosmic acceleration.

Now, because of the observed late-time cosmic coincidence, $R_{\Lambda_{obs.}}\approx R_{(c)}$, we get from equation \textcolor[rgb]{0.00,0.00,1.00}{(\ref{a65})}

\begin{eqnarray}
\mid \delta \mid \approx 0.6
\label{a66}
\end{eqnarray}

A small discrepancy in the value of $\delta$ in contrast to \textcolor[rgb]{0.00,0.00,1.00}{\cite{b77}}  may be attributed to the very small discrepancy in the exactness of the equality of $R_{\Lambda_{obs.}}$ and  $R_{(c)}$ \textcolor[rgb]{0.00,0.00,1.00}{\cite{b3}}. Thus, it seems that in an alternative gravity theories, an approximate ratio of the order of one of the present comparable energy densities is replaced by the necessity to constrain phenomenologically the model parameter corresponding to the dark sectors with an appropriate strength to fit the observational data.
Clearly, this is a simple analysis which could be used to constrain the shifted $f(R)$ model parameter with the realization of its consistency in different conformal frames.

In this way, different direct evidences from the observations at different scales can precisely constrain the shifted $f(R)$ model parameter $\delta$ along with its conformal analogue and it is also realized that $\delta$ can acts as a diagnostic tool to distinguish the two frames and also to provide the precise observational constraints via the combined analysis.

\section{\label{8}Summary and Discussions}
Let us then recapitulate the present work briefly. The study of precision cosmology is not possible in GR frame work until we completely know the physical system, but it provides a way to model the deviations in Einstein-Hilbert GR action. Deviations from Einstein GR theory are indeed predicted mostly in various extra-dimensional theories.  Contrasting alternatives to Einstein GR are actually useful to understand precisely which features of the theory have been tested in a particular experiment, and also to suggest new experiments probing different features. Such study can be modelled in the $f(R)$ gravity framework. Therefore, we study the deviations in GR in the form $R^{1+\delta}$ with $\delta$ being the  dimensionless physical observable quantity. We focus on the study of null and non-null geodesics at local scales and explore it under conformal $f(R)$ theory. Our discussion shows that $f(R)$ conformal transformation of spacetime metric ($g_{\mu\nu}$) leaves the null geodesic unchanged (Section III). On the other hand, the effect of such transformation for non-null geodesics produces an additional force on the unit test mass which can be traced under the Newtonian limit  (see equation \textcolor[rgb]{0.00,0.00,1.00}{(\ref{a36})} or \textcolor[rgb]{0.00,0.00,1.00}{(\ref{a37})}). We discuss the extent of this additional force in $f(R)$ Schwarzschild background and investigate its relativistic effect on the perihelion advance of Mercury orbit by obtaining the expressions for the $f(R)$ conformally related orbital precession of Mercury. We also explore some features of the $f(R)$ model for small deviations in Figs. (1), (2) and (3), whereas in Figs. (4), (5) and (8) we investigate the physical observations of the system confronted with different geodesics along with its conformal analogue and explore the precise deviations at different scales.
The main motivation to explore the deviation $\delta$ in the narrowest value comes from the recent combined investigation of
galactic dynamics for the power-law $f(R)$ model \textcolor[rgb]{0.00,0.00,1.00}{\cite{b33,b34,b36}} as well as from \textcolor[rgb]{0.00,0.00,1.00}{\cite{b27,b37}}.

The extra force can be screened in the high density environment as it directly depends on $\delta$ with $\delta \ll 1$  (see equation \textcolor[rgb]{0.00,0.00,1.00}{(\ref{aa37})}) and the observational results can be traced with smaller $\delta$ (Fig. 8). Such deviation can be envisaged as a distinguishing factor since $\delta$ can be specified as $v_{orbital}^2/c^2$ as well as also for the conformal investigations of different frames under different physical observations at local scale.

The phenomenological study of different  geodesics, light deflection angle for null geodesics and perihelion advance of mercury orbit for non-null geodesics (under an idealistic consideration\footnote{\textcolor[rgb]{1.00,0.00,0.00}{It is often stated in literature that the relativistic perihelion advance of Mercury is really only a test of the vacuum Schwarzschild solution, as  the general relativistic effects can be derived simply from such metrics.}} along with its conformal analogue at local scales imposes the strict constraints on  $\delta$ to be approximately $\mathcal{O}(10^{-6})$ with a difference of unity in order at galactic and planetary scales respectively (see Figs. 4, 5 and 8). Such constraints are strict in the sense that for orbital motions at local scales, $\delta$ can be physically represented as $v_{orbital}^2/c^2$ with known $v_{orbital}$ through observations. Our recent analysis of lensing profile of galaxies with $\delta \approx \mathcal{O}(10^{-6})$ for the dark matter $f(R)$ model may also, in principle, diagnose the present model from the other  galactic standard dark matter model (pseudo-isothermal sphere model etc.).  The analysis of perihelion advance of mercury orbit as contrast to other authors \textcolor[rgb]{0.00,0.00,1.00}{\cite{b37,b037,b74}} shows the new constraints within bare $f(R)$ framework to be about $\mathcal{O}(10^{-7})$ . Here, we do not consider the frame-dragging effect on the metric. Since, different missions (surveys) have reported an uncertainty in the GR result for the Mercury orbit \textcolor[rgb]{0.00,0.00,1.00}{\cite{b62,b63,b64,b65,b66,b74}}, so $f(R)$ gravity theory may be a potential candidate in this regard. Thus, the present model preserves the same form for the explanation of local scale dynamics including the planetary scales and hence an unique $f(R)$ at local scale.

We further analyse the constraint on the deviation parameter, $\delta$  via using the bare scalar self-interaction $f(R)$ potential to provide a null test of dark energy. In this direction, we discuss the Einstein frame $f(R)$ potential and explore the cosmic late-time constraint on the model parameter to be $\mid\delta\mid\approx 0.6$ which is a close agreement with the encouraging results explored in Jordan frame by Capozziello et al. \textcolor[rgb]{0.00,0.00,1.00}{\cite{b77}}.
Our analysis provides a powerful tool to obtain the precise constraints on the shifted parameter modelled in $f(R)$ gravity at different scales.

Such smaller values for $\delta$ are also consistent with cosmological constraints coming from primordial nucleosynthesis \textcolor[rgb]{0.00,0.00,1.00}{\cite{b27,b78,b79}}.

Currently all operating projects for the detection of gravitational waves, including LIGO, Virgo and LISA are based on geodesic deviation equation, so we will also extend this work in future for the study of binary systems and gravitational wave (GW) analysis and will look for the physical quantity for $\delta$ to directly diagnose the model with  GW data in order to investigate uncertainty in the Einstein's GR theory.

The present status of the custered (DM) and unclustered (DE) dark sector hypothesis and the modified
gravity alternatives, deduce that there is a lack of definitive convincing arguments in favour of any of the two concepts or particular theories. Basically, all the gravity
models have some successful predictions but also problematic comparisons with some observations and experiments.
Therefore, it seems that new observations are necessary, especially on the scale of stars and planetary systems.

As an important comment, because of the highly different values of the shifted parameter $\delta$ of $f(R)$ model at different scales (basically at local scale and at late-time cosmic scales), it may not be a suitable model in its present form to study the unified approach to dark matter, dark energy and inflation. The possible investigation of precise modification in this direction (at early time epoch) is also explored by the authors \textcolor[rgb]{0.00,0.00,1.00}{\cite{b80}}.  Although, the investigated model seems to be consistent for the study of clustered dark matter like dynamics at local scales and therefore would bridge the gap among the gravitational anomalies ranging from stellar to
galactic scales.
In this way, different direct evidence from the observations at different scales can precisely constrain the shifted $f(R)$ model parameter $\delta$ along with its conformal analogue.
Our findings are also in an close agreement with the recently explored values by us at galactic scales and also seems to provide a new phenomenological constraint for the study of perihelion advance of Mercury orbit as it differs from the values explored by different authors.

\section*{Acknowledgments}
Authors thank  IUCAA, Pune, for providing  the facilities  with which a part of the present work was completed under the associateship programme.  VKS  also thanks  Swagat S. Mishra, B. K. Yadav and A. K. Sharma for fruitful discussions. He  thanks  Varun Sahni for several comments  and support during visits to IUCAA.\\


\end{document}